# Observation-Specific Uncertainty Modeling for Latent-Variable Regression in High-Dimensional Analytical Data

Carl Emil Eskildsen[1,2,3‡], Georgios Papadopoulos[2,3,4], Kevin A. Giraldo[2,3], Mauricio Barahona[5], Molly M. Stevens[1,2,3,6,7,8‡]

[1]Department of Bioengineering, Imperial College London, London SW7 2AZ, United Kingdom

[2]Department of Physiology, Anatomy and Genetics, University of Oxford, Oxford OX1 3PT, United Kingdom

[3]Kavli Institute for Nanoscience Discovery, University of Oxford, Oxford OX1 3QU, United Kingdom

[4]Department of Computing, Imperial College London, London SW7 2AZ, United Kingdom

[5]Department of Mathematics, Imperial College London, London SW7 2AZ, United Kingdom

[6]Department of Materials, Imperial College London, London SW7 2AZ, United Kingdom

[7]Institute of Biomedical Engineering, Imperial College London, London SW7 2AZ, United Kingdom

[8]Department of Engineering Science, University of Oxford, Oxford OX1 3PJ, United Kingdom

**[‡]Correspondence**

Carl Emil Eskildsen,

Email: carl.eskildsen@dpag.ox.ac.uk

Molly M. Stevens,

Email: molly.stevens@dpag.ox.ac.uk

## ABSTRACT

Across many fields, including medical diagnostics, process monitoring, and quality control, decisions often rely on point predictions for single samples. Observation-specific uncertainty is needed to assess the risk of decision error when predictions are compared with thresholds. Latent-variable regression is widely used for such predictions, but observation-specific uncertainty remains difficult to estimate. Traditional methods assume unbiased predictions, attributing prediction uncertainty solely to variance. However, dimensionality reduction can introduce observation-specific bias, violating this assumption. We introduce a framework using Gaussian processes trained on latent scores and prediction errors to estimate observation-specific bias and error variance, separating systematic error from epistemic and aleatoric variance. In Raman measurements of a fructose–sucrose model system, estimated prediction-error distributions agreed with empirical distributions; bias correction reduced mean squared error, and decision limits incorporating bias and variance improved analyte detection over variance-only limits. An aqueous-solubility application using molecular descriptors demonstrated the framework's generalisation beyond spectroscopy.

## Introduction

Accurate observation-specific uncertainty estimates are crucial in decision-making contexts such as medical diagnostics,[1,2] process monitoring,[3-5] and quality control.[6] In medical diagnostics, for example, a predicted biomarker concentration slightly below a clinical threshold may seem safe at first glance, yet consideration of its uncertainty could significantly alter the decision about clinical intervention.[7] In process monitoring applications, Deming[8] also emphasised that process adjustments based solely on point estimates (i.e., without acknowledging prediction uncertainty) can inadvertently degrade process performance, a practice he termed tampering. Likewise, in quality control (e.g., determining whether a sample is adulterated), decisions rely both on point predictions and the distribution of comparable non-adulterated samples.[9] Ignoring uncertainties associated with individual predictions in these scenarios can lead to suboptimal or even hazardous outcomes. Despite these well-documented pitfalls, observation-specific uncertainty is still rarely considered in the analysis and interpretation of predictive models.

Latent variable (LV)-based regression methods, such as partial least squares (PLS),[10] are widely adopted across analytical disciplines because of their ability to reliably compress high-dimensional data, providing both regularised regression estimates and interpretable low-dimensional representations.[4,11,12] This relevance extends beyond traditional chemometrics to omics-based analyses, where high-dimensional prediction and classification are also common[13,14]. Despite their importance, observation-specific uncertainty estimates remain challenging for LV-based regression models.[15] By compressing high-dimensional data into a lower-dimensional latent space, LV-based methods introduce systematic deviations that can result in observation-specific prediction bias.[15] Recent work has shown that such bias cannot generally be ignored and, in specific cases, can be reduced through constrained regression.[15,16] Observation-specific uncertainty in LV-based regression must account for both systematic prediction error and random variation.

Traditionally, estimation of observation-specific uncertainties for LV-based predictions has been rooted in ordinary least squares (OLS)-derived methods.[3,17,18] In such approaches, uncertainty estimates ultimately require an estimate of random error, often obtained from residual variance or mean-squared-error (MSE)-type quantities.[3] This is appropriate when residuals are unbiased, because the error term can be treated as zero-mean and its variance reflects random scatter around the prediction. In LV-based regression, however, residuals can contain systematic, observation-specific prediction bias.[15,16,17] If an MSE-like quantity is then used as the random-error estimate, it includes both random scatter and the squared contribution of systematic error. The resulting uncertainty estimate can therefore become inflated, because systematic error is represented as additional variance rather than as a non-zero expected error.

LV regression does not admit a simple general closed-form expression for observation-specific uncertainty, which has motivated a broad set of approximate and empirical approaches, including analytical error propagation,[19,20] sample-specific standard errors[21] and prediction intervals for PLS,[22] residual resampling, cross-validation, error-in-variables formulations,[19] distribution-free predictive inference,[23] latent-space uncertainty calibration,[24] and local linearization of PLS prediction-error variance.[22,25] This literature reflects a broader recognition that LV-based predictions should be accompanied by uncertainty estimates that account for the underlying error structure and data-processing workflow.[26] However, these approaches generally quantify uncertainty through standard errors, interval widths, or empirical coverage, rather than explicitly separating observation-specific expected prediction error from residual random variance. When systematic error is present, this lack of separation can make uncertainty estimates incomplete or overly conservative for decision-making in an LV context. Here, we aim to make this separation directly in latent space, estimating systematic prediction error for bias correction and variance terms for observation-specific probabilistic decision-making.

LV-based methods such as PLS define a continuous mapping from a high-dimensional space of observed/explanatory variables, $\mathbf{X} \in \mathbb{R}^{N \times M}$ to a low-dimensional space of latent scores, $\mathbf{T} \in \mathbb{R}^{N \times A}$

$$\mathbf{X} \mapsto \mathbf{T} = \mathbf{XW} \quad (1)$$

where $\mathbf{W} \in \mathbb{R}^{M \times A}$ are weights; $N$ is the number of observations; $M$ is the number of original variables; and $A$ is the number of LVs (i.e., the dimension of the latent space).[27] Note that $A < \min\{N, M\}$, thus ensuring dimensionality reduction. Because LV-based models preserve a covariance structure, samples that map close together in the latent-space are expected to share similar prediction errors—both in terms of their expected value (i.e. negative bias) and their variance. In other words, the error structure has a local structure that varies continuously across the latent manifold. Recognizing this locality is key: it suggests that the score values of an observation can inform the tailored estimation of its prediction-error distribution.

To operationalize this idea, we feed latent scores as input to a Gaussian process (GP) regression pipeline,[28] with the corresponding prediction errors as output. A first GP—hereafter referred to as the error GP—learns the spatially-varying expected prediction error. Its posterior variance (i.e., the uncertainty about the expected prediction error) quantifies epistemic uncertainty, reflecting limited training data.[28] A second heteroscedastic GP—hereafter referred to as the variance GP—estimates the aleatoric variance, reflecting the irreducible uncertainty that may differ across the latent space and introduce heteroscedasticity.[29,30] The resulting framework thus decomposes prediction uncertainty into interpretable components and yields robust, observation-specific uncertainty estimates that accurately reflect the complex data projection intrinsic to LV-based analyses.

By embedding GP regression within the PLS framework, our method provides: (1) a unified solution for systematically correcting prediction biases while simultaneously providing reliable variance estimates at the observation level; (2) observation-tailored detection limits robust to interferents; and (3) analytical insights into latent-space error structures. This improved uncertainty quantification framework for LV-based models can thus facilitate informed decision-making across diverse analytical domains.

## Results

We demonstrate our framework on a two-constituent model system of fructose (analyte) and sucrose (interferent; Fig. 1A). Samples were prepared as detailed in the Supplementary Information (SI), Supplementary Note 1, Sample preparation, yielding known analyte responses, $\mathbf{y} \in \mathbb{R}^N$. Raman spectra were recorded (Supplementary Note 1, Raman spectroscopy) and preprocessed (Supplementary Note 1, Spectral preprocessing) to form the explanatory variables, $\mathbf{X}$ (Fig. 1B).

Next, we calibrated a PLS regression model relating $\mathbf{X}$ to $\mathbf{y}$ (Supplementary Note 1, Partial least squares regression; Fig. 1C). For each observation $i$, this procedure yields the latent scores, $\mathbf{t}_i \in \mathbb{R}^A$, the prediction, $\hat{y}_i$, and the prediction error, $e_i$,

$$e_i = y_i - \hat{y}_i \,. \quad (2)$$

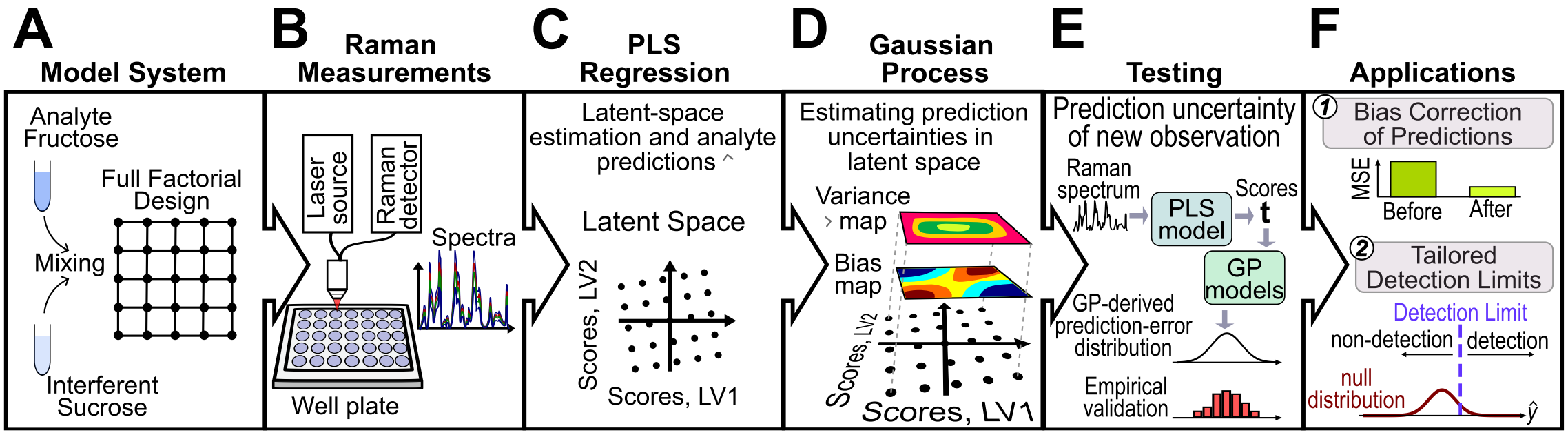

**Fig. 1 | Study overview.** **A** Preparation of samples in a two-constituent system (fructose analyte + sucrose interferent). **B** Acquisition of Raman spectra. **C** Partial least squares (PLS) regression calibrated on the calibration set, relating Raman measurements to analyte concentrations in a latent space. **D** Training of Gaussian process (GP) models on validation-set PLS scores and corresponding prediction errors to learn spatially varying systematic prediction error (i.e., bias) and error variance across the latent space. The coloured surfaces schematically represent these quantities. **E** Testing of GP-derived prediction-error distributions against empirical distributions estimated from test-set observations. **F** Application of GP-derived uncertainties: (1) bias correction reduces mean squared error (MSE) of analyte predictions, and (2) tailored detection limits enable reliable analyte presence/absence decisions. The purple vertical line denotes the observation-specific detection limit. Latent variable (LV); PLS score vector ($\mathbf{t}$); Analyte prediction ($\hat{y}$).

We decomposed this error into: (1) an expected prediction error, $\mathbb{E}[e_i]$; (2) an epistemic variance, $\sigma^2_{\mathrm{epi}_i}$; and (3) an aleatoric variance, $\sigma^2_{\mathrm{ale}_i}$, each of which may depend on latent-space location.[15,16] For details, see Methods, Prediction-error decomposition.

We then trained two GP models on the latent scores and prediction errors (Methods, Gaussian-process models; Fig. 1D). The error GP estimated the expected prediction error (Eq. 11) and epistemic variance (Eq. 12) across the latent space, and the variance GP estimated the aleatoric variance from debiased residuals (Eqs. 13-18). Together, they provide, for any new observation, a bias-corrected prediction and an observation-specific uncertainty estimate. For details, see Methods Gaussian-process models.

We tested the GP-derived uncertainty distributions against empirical test samples (Methods, Model testing; Fig. 1E) and assessed their impact on prediction and analyte detection (Methods, Gaussian-process-corrected predictions and Analyte detection; Fig. 1F).

## Experimental design of the Raman model system

A total of 37 analyte-interferent (fructose-sucrose) samples with different compositions were prepared (Fig. 2). For each sample composition, 32 wells were prepared in a 96-well plate, and three Raman spectra were recorded per well, yielding 96 observations.

Of the 37 sample compositions, 25 followed a $5^2$ full-factorial analyte-interferent design and were used for both model development and method testing (Fig. 2). Wells were randomly assigned to calibration (2 wells per sample; 150 observations total), validation (5 wells per sample; 375 observations total), or test (25 wells per sample; 1,875 observations total), while keeping the three repeated measurements from each well together.

The remaining 12 sample compositions represented additional analyte–interferent combinations outside the full-factorial design and were used only for testing (Fig. 2). For each of these sample compositions, 25 wells were randomly selected for inclusion in the test set so that all test-set samples contributed the same number of observations, adding 900 observations for a total test set of 2,775 observations.

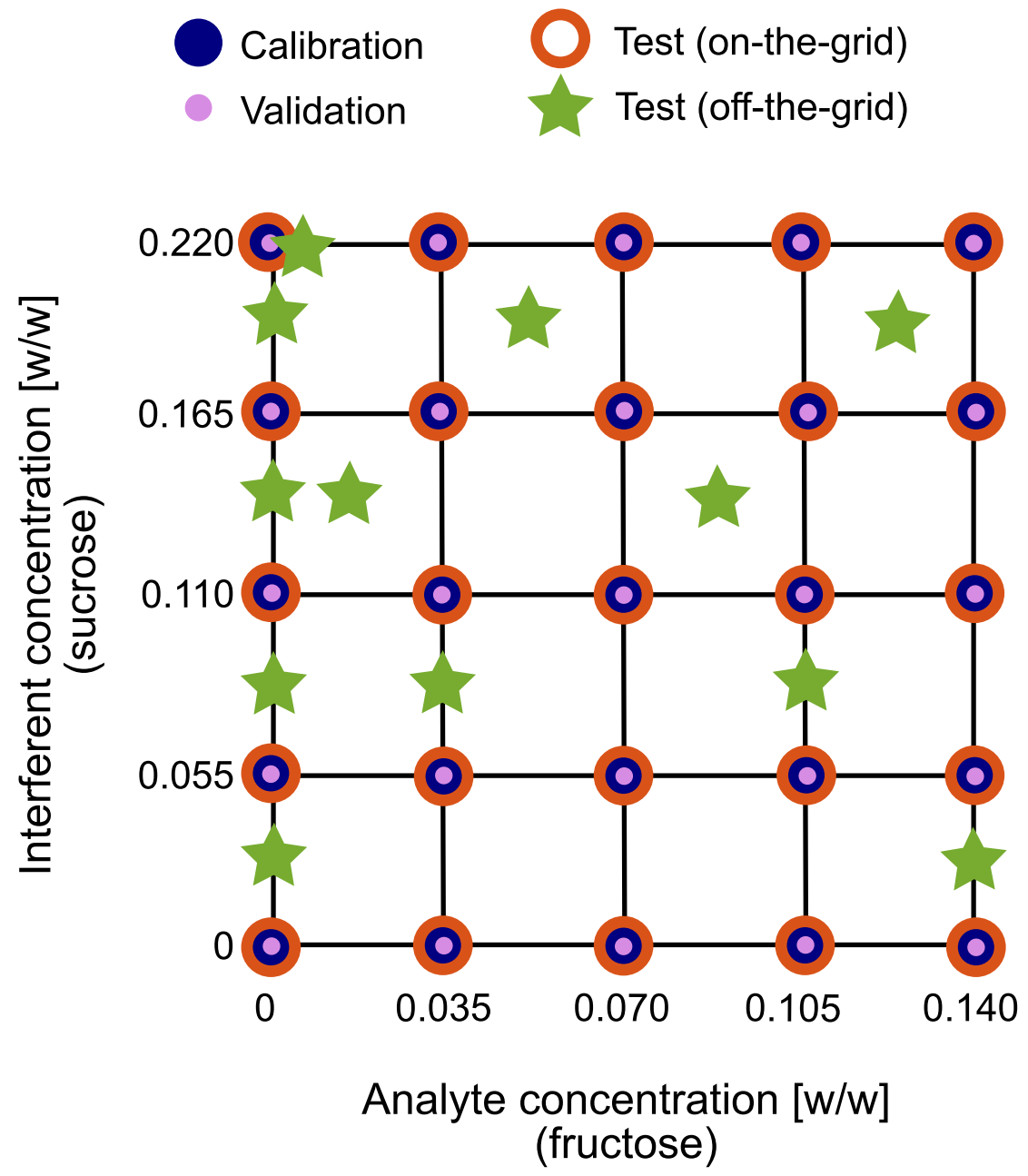

**Fig. 2 | Experimental design.** For each of the 37 sample compositions, 32 wells were prepared, and three Raman spectra were recorded from every well. Of these, 25 sample compositions followed a full-factorial grid of analyte (fructose, 0–0.140 w/w) and interferent (sucrose, 0–0.220 w/w) concentrations. Wells from these 25 sample compositions were randomly assigned to calibration (2 wells per sample), validation (5 wells per sample), and test (25 wells per sample) subsets, while keeping the three repeated measurements from the same well together. The remaining 12 sample compositions represented additional analyte–interferent combinations not included in the full-factorial design; for these, 25 wells per sample were randomly selected for inclusion in the test set. Symbols indicate dataset membership at each composition: blue circles represent calibration data, purple circles validation data, orange rings on-the-grid test data, and green stars off-the-grid test data.

## Analyte quantification

A PLS model was fitted to the 150 calibration observations. Leave-one-sample-out cross-validation (CV) determined the optimal number of LVs, and the final model was fitted to all calibration observations.[31] Preprocessed and mean-centred explanatory variables, preprocessed pure signals of analyte and interferent, and the corresponding PLS loadings are presented in Supplementary Fig. 1.

Fig. 3 summarises the calibration results. Fig. 3A displays the leave-one-sample-out CV MSE as a function of the number of LVs and suggests a rank-two structure, consistent with the presence of one analyte and one interferent.

Fig. 3B shows the analyte predictions, $\hat{\mathbf{y}} \in \mathbb{R}^N$, plotted against the measured analyte levels, $\mathbf{y}$, for the different interferent levels. A systematic pattern is observed: at low-analyte levels, high-interferent observations tend to be underestimated (positive residuals), while low-interferent observations are overestimated (negative residuals). This pattern is reversed at high-analyte levels. Supplementary Fig. 2 shows the residuals $\mathbf{e} = \mathbf{y} - \hat{\mathbf{y}}$ plotted against $\mathbf{y}$. These results indicate that interferent levels systematically bias the PLS predictions, with direction and magnitude depending on both analyte and interferent concentrations.

Fig. 3C presents the PLS scores of LV1 and LV2 for all calibration observations. LV1 captures increasing analyte levels (Supplementary Fig. 3), whereas LV2 reflects increasing interferent levels, as indicated by the colour gradient in Fig 3C. Thus, the 2D latent space aligns with the experimental design in Fig. 2, and provides the score representation used as input to the GP pipeline, with prediction errors as outputs.

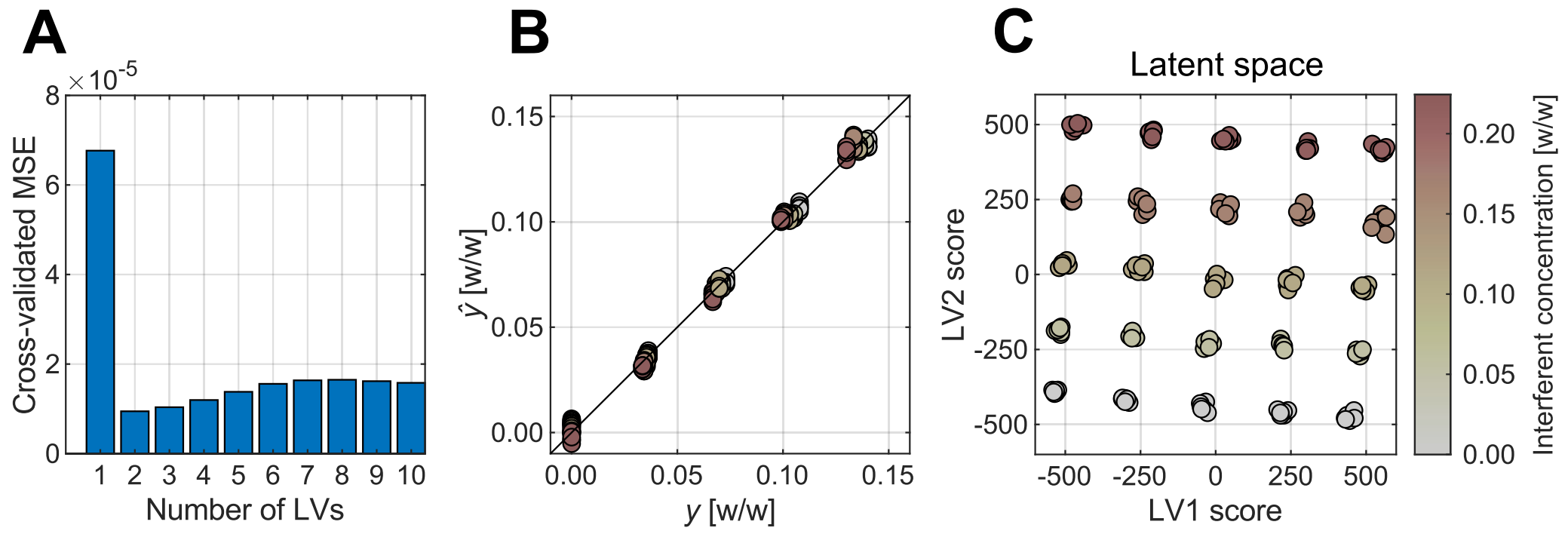

**Fig. 3 | Calibration of the partial least squares (PLS) regression model. A** Cross-validated mean squared error (MSE) as a function of the number of latent variables (LVs). **B** Predicted analyte (fructose) concentration ($\hat{y}$) versus measured analyte ($y$) concentration. The black line denotes the identity line, $\hat{y} = y$. **C** Calibration-set scores for latent variables 1 and 2, defining the latent space used for subsequent Gaussian-process modelling of prediction errors. In **B** and **C**, observations are coloured by interferent (sucrose) concentration using the colour scale shown in **C**.

## Estimation of prediction-error distributions

The fitted PLS model was used to estimate PLS scores and prediction errors of the validation set (375 observations). The GP pipeline was then trained using the PLS scores and prediction errors of the validation set. Fig. 4A presents a contour map of the GP-derived expected prediction error overlaid on the PLS score plot. We find a distinct pattern: observations with either low scores on both LV1 and LV2 or high scores on both LV1 and LV2 exhibit negative expected prediction errors, whereas observations with low LV1 and high LV2 scores, or the reverse, show positive expected prediction errors. Expected prediction errors are also reduced for observations located near the center of the model latent space.

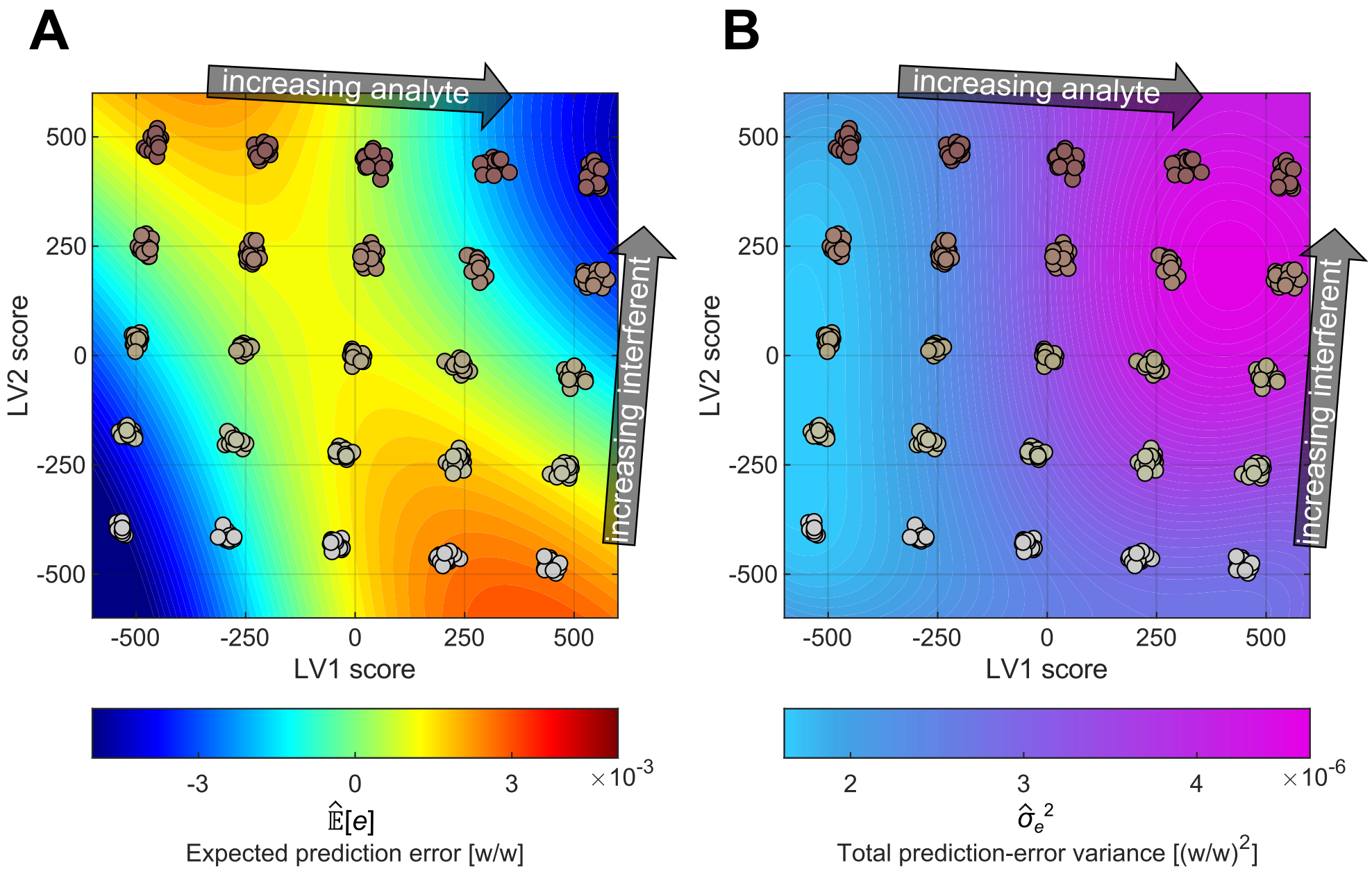

**Fig. 4 | Gaussian-process estimates across the partial least squares latent space. A** Expected prediction error ($\hat{\mathbb{E}}[e]$). **B** Total prediction-error variance ($\hat{\sigma}_e^2$), calculated as the sum of epistemic and aleatoric variance components. Background contours show the Gaussian-process estimates, and circles indicate validation observations coloured by interferent (sucrose) concentration from grey (low) to brown (high), consistent with Fig. 3C. Arrows indicate the directions of increasing analyte and interferent concentration. LV, latent variable.

Fig. 4B displays the GP-derived total prediction-error variance, i.e. the sum of the epistemic and aleatoric components (Eq. 9, Methods, Prediction-error decomposition), overlaid on the PLS score plot. Because the variance GP models heteroscedastic noise, the estimated variance adapts to local conditions across the latent space. The total prediction-error variance is lowest in the lower-left region of the score plot and increases toward the upper-right region, corresponding to increasing analyte and interferent levels.

Decomposition of the total variance into epistemic and aleatoric components (Fig. 5) shows that this spatial pattern is driven primarily by the aleatoric term. The epistemic variance is small overall and increases only modestly toward the boundaries of the latent space, consistent with weaker local support for the GP estimate in those regions. By contrast, the aleatoric variance varies systematically across the score plot, indicating that the dominant source of uncertainty in this system is composition-dependent heteroscedasticity rather than lack of local model support.

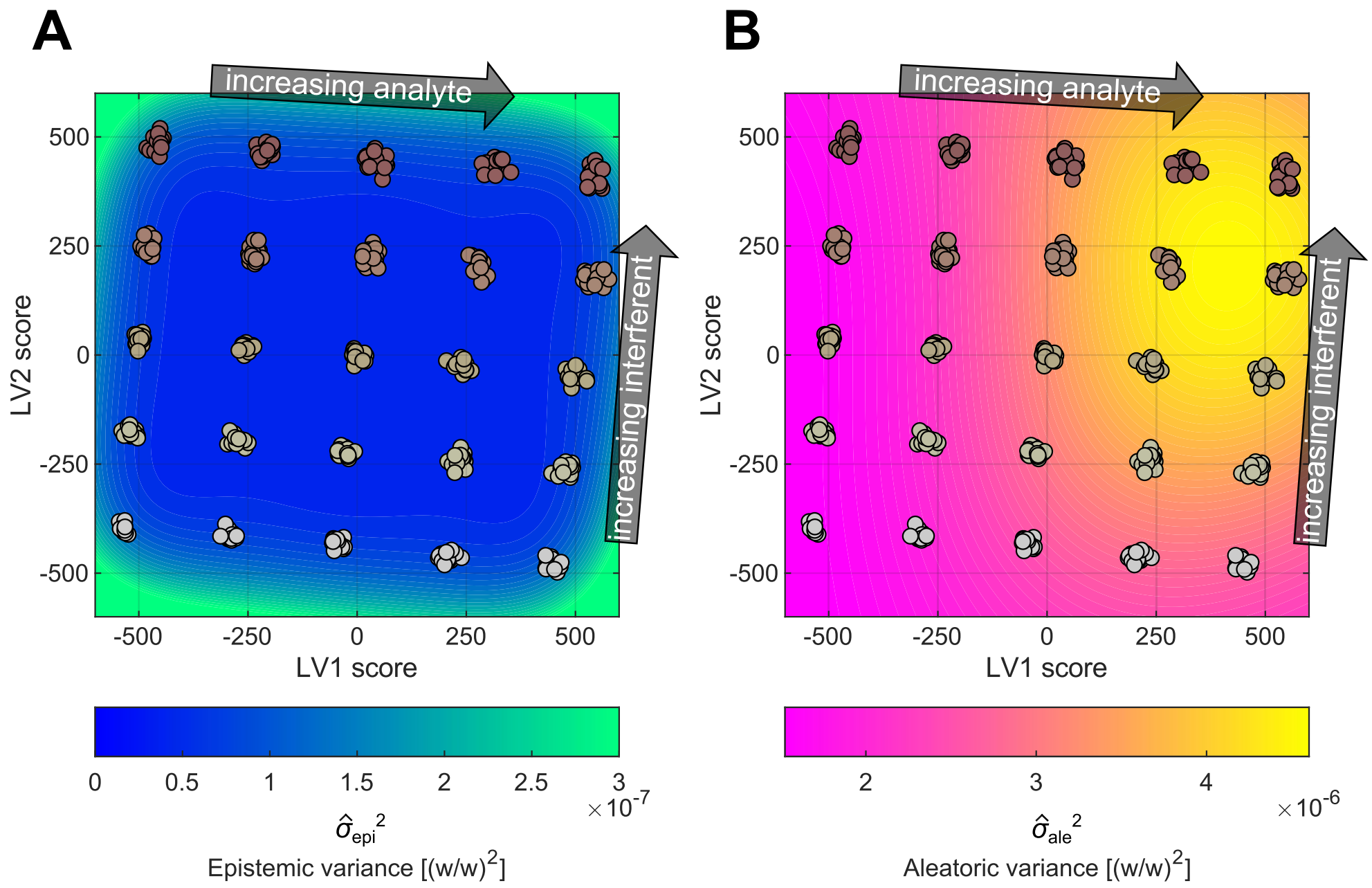

**Fig. 5 | Gaussian-process decomposition of prediction-error variance across the partial least squares latent space.** **A** Estimated epistemic variance ($\hat{\sigma}^2_{\mathrm{epi}}$). **B** Estimated aleatoric variance ($\hat{\sigma}^2_{\mathrm{ale}}$). Together, these components sum to the total prediction-error variance shown in Fig. 4B. Background contours show the Gaussian-process estimates, and circles indicate validation observations coloured by interferent (sucrose) concentration from grey (low) to brown (high), consistent with Fig. 3C. LV, latent variable.

## Evaluation of prediction-error distributions against the test sets

The PLS model and GP models were applied to the held-out test sets to provide an independent assessment of GP-derived prediction-error distributions. We first used the on-the-grid test set measurements for a structured assessment of how prediction-error distributions vary with analyte and interferent levels. Fig. 6 presents histograms of empirical prediction errors overlaid with GP-derived error distributions, organised from samples containing no analyte (Fig. 6A) to the highest analyte level (Fig. 6E). For each test sample, LV scores were averaged across the 75 observations, and the resulting mean score was used as input to the GP models. The overlaid GP-derived distributions therefore describe how prediction errors vary around the estimated mean error at that representative latent-space location, i.e. through the aleatoric variance, whereas the epistemic component—reflecting uncertainty in the GP estimate of that mean error—was excluded from this comparison.

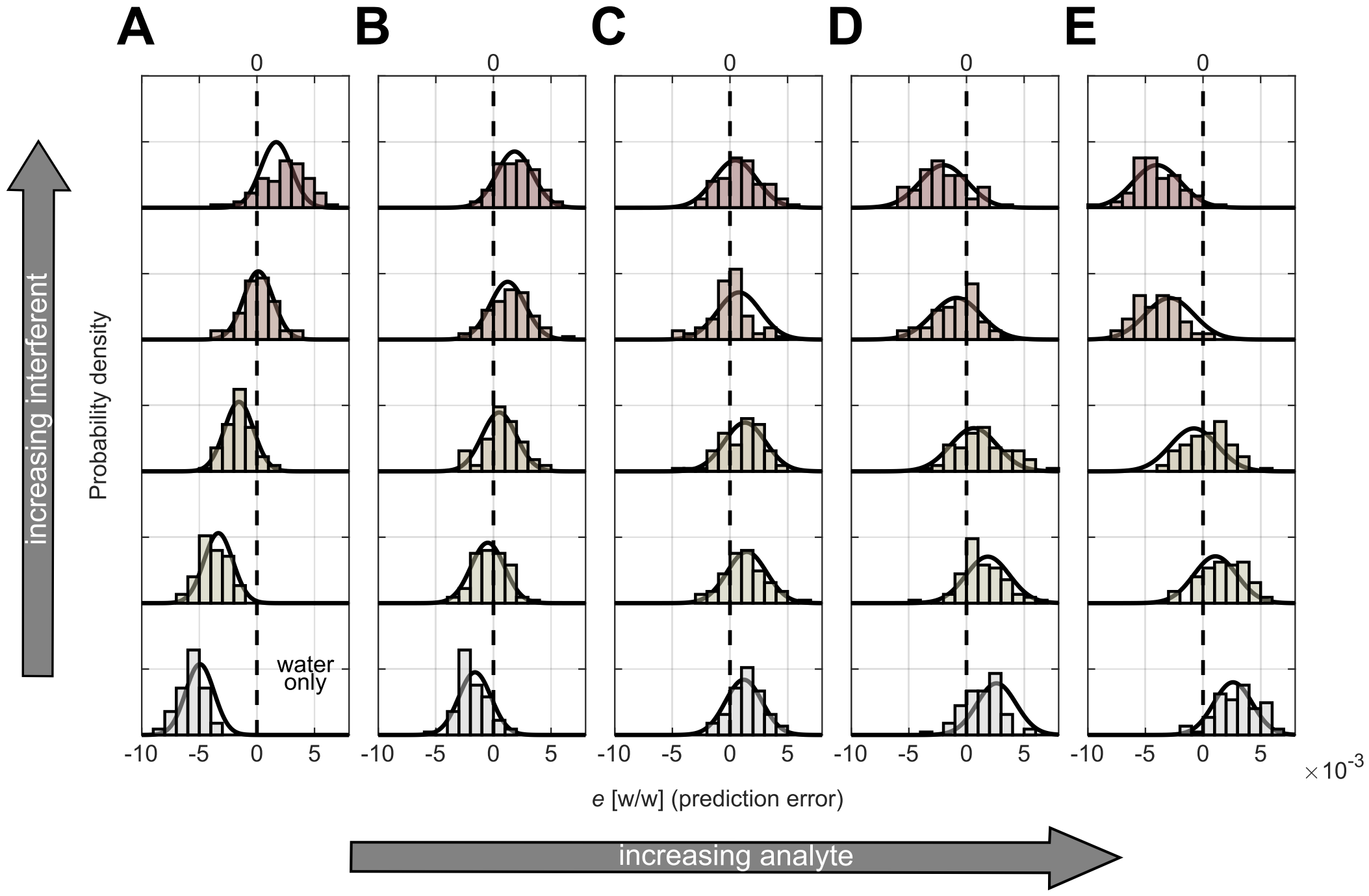

**Fig. 6 | Empirical and Gaussian-process-derived prediction-error distributions for the on-the-grid test set.** **A–E** Samples with increasing analyte (fructose) concentration, from **A**, containing no analyte, to **E**, containing the highest analyte concentration. Within each panel, vertically offset rows correspond to increasing interferent (sucrose) concentration from bottom to top; the bottom row of **A** represents water only. Histograms show empirical prediction errors from 75 observations per sample, and black curves show Gaussian-process-derived normal distributions parameterised by the estimated expected prediction error and aleatoric variance; epistemic variance is excluded. Histogram colours indicate interferent concentration from grey (low) to brown (high), consistent with Fig. 3C. Dashed vertical lines indicate zero prediction error.

The agreement between empirical and GP-derived error distributions shows that the GP captures observation-specific prediction uncertainties. The results in Fig. 6 reveal spatially structured residuals inherent to the PLS framework, consistent with Fig. 4A. At low analyte levels (Fig. 6A–B), high interferent levels produce positive residuals, indicating underestimation, whereas low interferent levels produce negative residuals, reflecting overestimation. This pattern reverses at high analyte levels (Fig. 6D–E). Near-mean analyte and interferent levels (Fig. 6C) minimise residuals, consistent with the mean-centred nature of the PLS model, which balances deviations across the latent space.

We also evaluated the framework on 12 off-the-grid test samples (Supplementary Fig. 4), representing previously unseen analyte–interferent combinations. The GP-derived error distributions remained accurate, demonstrating the robustness of the framework.

Fig. 7 compares GP-derived expected prediction errors and aleatoric variances with their empirical counterparts across all test samples. Fig. 7A shows that the GP-derived expected prediction error closely tracks the empirical mean prediction error across both test sets, confirming that the error GP captures the spatially structured pattern in expected prediction error identified in Fig. 4A. Fig. 7B shows that the GP-derived aleatoric variance captures the general magnitude and spread of the empirical sample variances. Although some scatter remains, because empirical variances estimated from limited observations are inherently noisy, the GP provides stable, spatially smoothed variance estimates suitable for constructing observation-specific prediction intervals.

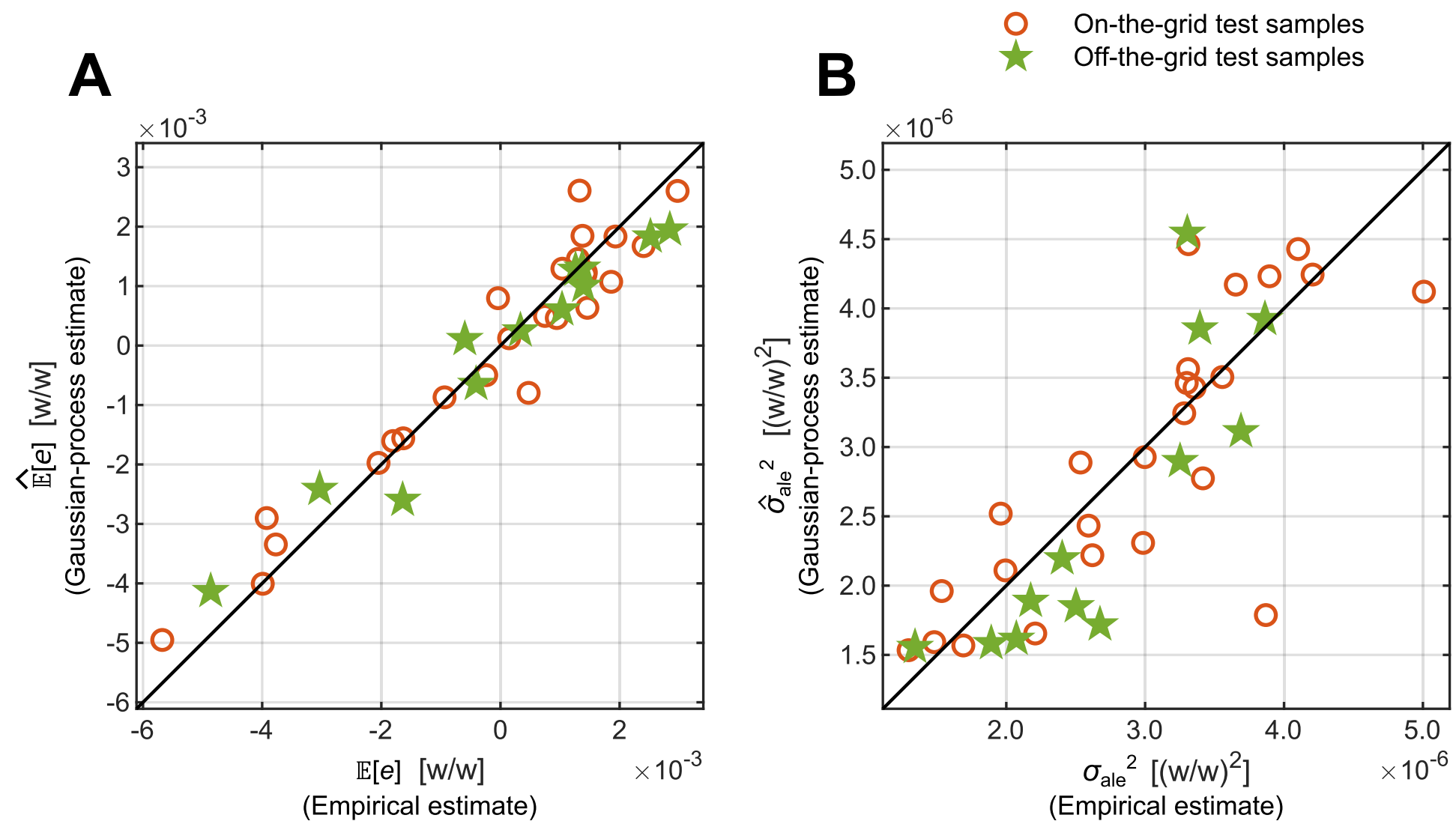

**Fig. 7 | Gaussian-process estimates compared with empirical prediction-error statistics. A** Gaussian-process estimate of expected prediction error ($\hat{\mathbb{E}}[e]$) versus the empirical mean prediction error ($\mathbb{E}[e]$). **B** Gaussian-process estimate of aleatoric variance ($\hat{\sigma}^2_{\mathrm{ale}}$) versus the empirical aleatoric variance ($\sigma^2_{\mathrm{ale}}$). Empirical expected prediction errors and aleatoric variances were calculated as the mean and sample variance, respectively, of prediction errors across 75 observations per test sample. Orange circles and green stars denote on-the-grid and off-the-grid test samples, respectively. Solid black lines indicate equality.

### The origin of the link between error structure and latent space

The spatially structured residuals in PLS predictions arise from the interplay between LV scores and their regression coefficients ($q$-values, Eq. 3, Methods, Partial least squares regression). The two-LV PLS model provides a single global linear approximation across the calibration range, so its predictions are most accurate near the calibration mean and can deviate systematically towards the extremes of the latent space.

LV1, governed by a relatively large positive $q$-value (Supplementary Table 1), dominates predictions by capturing analyte-related variance in $\mathbf{X}$ (Supplementary Fig. 5A). Yet, spectral overlap between the analyte and interferent (Supplementary Fig. 1) causes PLS to conflate shared variance into LV1: at fixed analyte levels, higher interferent levels inflate LV1 scores and lower interferent levels deflate them, producing analyte over- and underestimation, respectively (Supplementary Fig. 5A–B). This LV1-driven distortion increases with analyte concentration. LV2, with its smaller negative $q$-value (Supplementary Table 1), partially offsets this by shifting predictions downward in high-interferent regions and upward in low-interferent regions, but its fixed corrective contribution becomes a compromise between correcting low analyte regions (low distortion) and high analyte regions (high distortion) (Supplementary Fig. 5B). As a result, LV2 over-corrects in low-analyte areas and under-corrects in high-analyte areas (Supplementary Fig. 5C), yielding spatially structured systematic residuals that diminish near the latent space center (Fig. 6C).

These systematic errors, which depend on the specific system under study and LV dimensionality, motivate observation-specific bias corrections such as our GP framework, which learns and exploits these spatial trends in the latent space.

## Gaussian-process-derived prediction error correction

The observation-specific GP-derived expected prediction errors were used to correct PLS predictions of all test samples (Eq. 19 in Methods, Gaussian-process-corrected predictions). Fig. 8A and B show the original and GP-corrected predictions, respectively, and Fig. 8C summarises the corresponding MSEs for the on-the-grid and off-the-grid test sets.

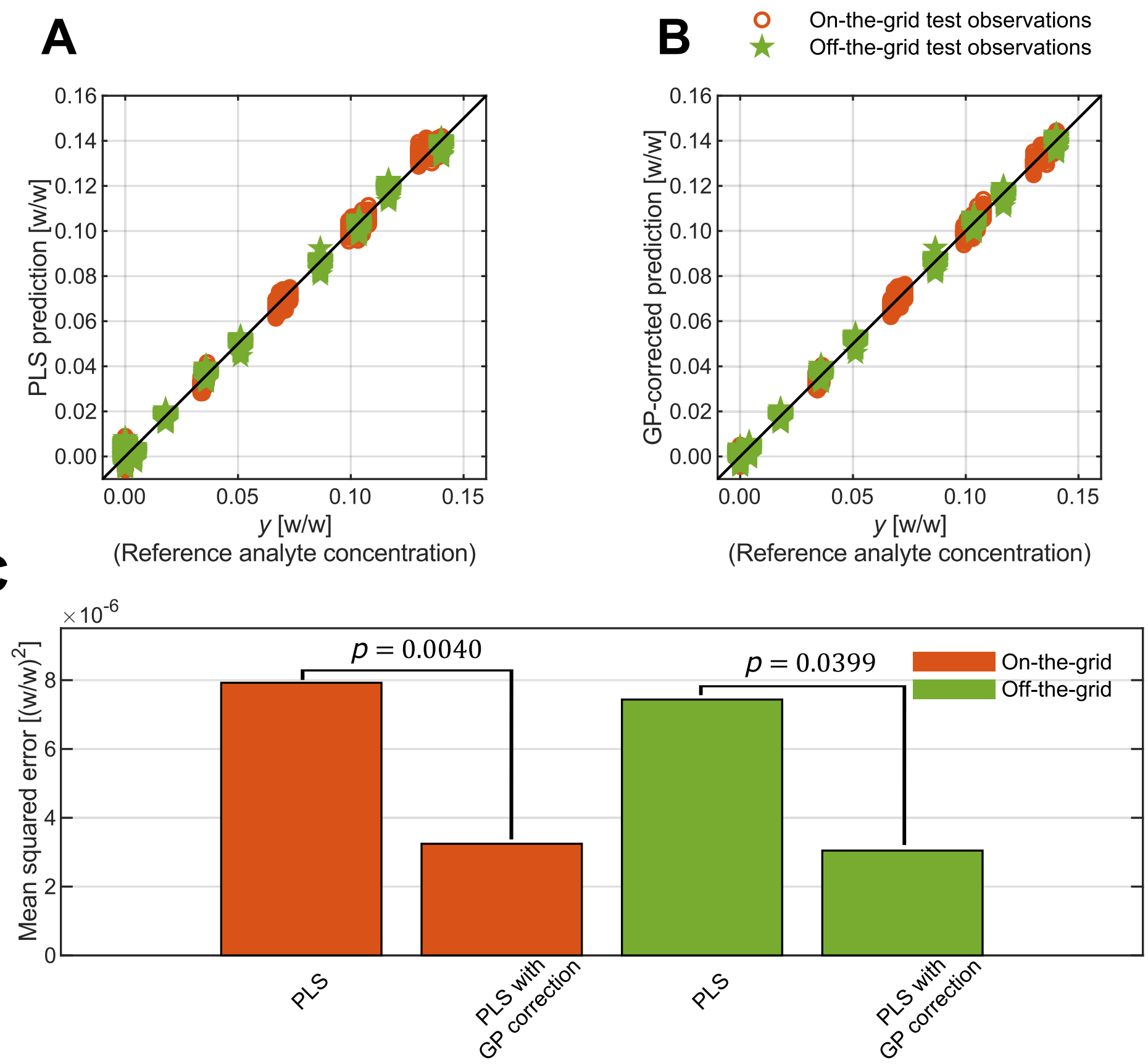

**Fig. 8 | Effect of Gaussian-process correction on partial least squares predictions. A** Uncorrected partial least squares (PLS) predictions ($\hat{y}$) versus reference analyte concentrations ($y$) for individual test observations. **B** Gaussian-process (GP)-corrected predictions ($\hat{y}^{\mathrm{corr}}$) versus reference analyte concentrations. Orange circles and green stars denote on-the-grid and off-the-grid test observations, respectively. Solid black lines indicate equality. **C** Mean squared error (MSE) before and after GP correction for the on-the-grid (orange) and off-the-grid (green) test sets. GP correction reduced MSE by 59.0% in both test sets. Correction was applied to individual observations, and statistical comparisons used paired MSEs averaged over the 75 observations within each mixture. Displayed $p$-values are from two-sided paired t-tests with $n = 25$ on-the-grid mixtures (24 degrees of freedom) and $n = 12$ off-the-grid mixtures (11 degrees of freedom). These $p$-values are interpreted as exploratory because the mixture-level MSE differences departed from normality (Methods, Statistical analyses).

For the on-the-grid test set, GP correction reduced the MSE from $7.92 \times 10^{-6}$ to $3.25 \times 10^{-6}$; for the off-the-grid test set, it reduced the MSE from $7.44 \times 10^{-6}$ to $3.05 \times 10^{-6}$, corresponding to a 59.0% reduction in both cases. Statistical comparisons used paired MSEs calculated within each test mixture before and after observation-specific correction, with 25 on-the-grid and 12 off-the-grid mixtures. Two-sided paired t-tests yielded $p = 0.0040$ and $p = 0.0399$, respectively. Wilcoxon signed-rank tests yielded $p = 0.0001$ and $p = 0.0068$, respectively. These tests are interpreted as exploratory because of departures from their distributional assumptions (Methods, Statistical analyses). GP correction reduced the overall MSE in both test sets.

### Analyte detection with observation-specific uncertainty

Determining whether an analyte is present in a sample typically involves comparing the analyte prediction with a null distribution, i.e., the distribution expected for blank observations containing no analyte. Predictions exceeding an upper confidence limit derived from this null distribution indicate analyte presence.[7] However, because the null distribution shifts with interferent concentrations (Fig. 6A), estimating it empirically for each interferent level quickly becomes impractical.[32,33]

To overcome this issue, we define a null-space within the latent space that captures the variation of blank samples (Eqs. 21–22 in Methods, Analyte detection). For a new sample, its LV scores are projected onto this null-space to obtain hypothetical blank scores, i.e. the scores the observation would have if no analyte were present (Eqs. 23–26). These hypothetical blank scores are then passed to the GP models to estimate the corresponding null distribution, including both the local expected prediction error and the local variance, from which the sample-specific detection limit is calculated (Eq. 27).

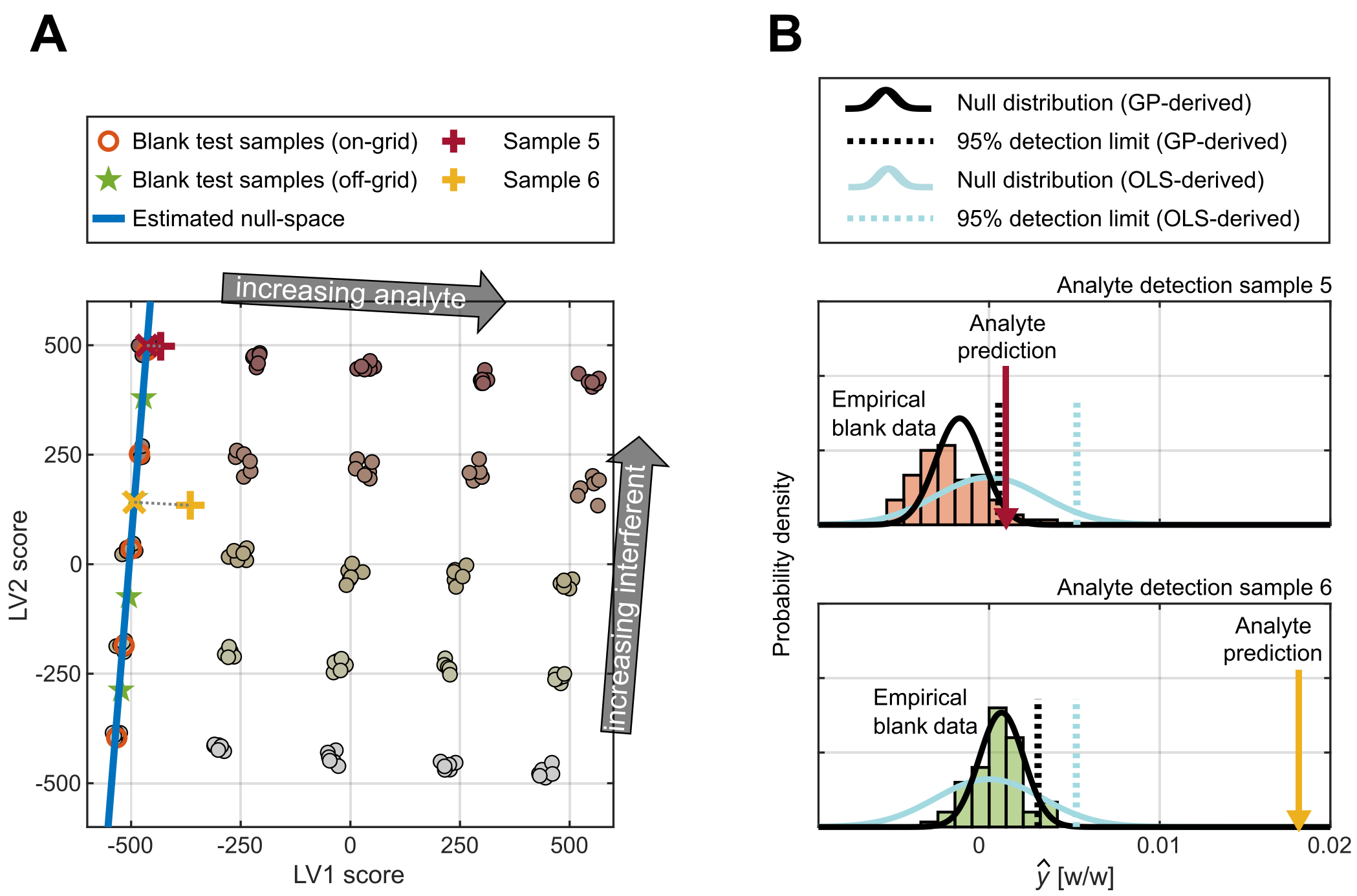

**Fig. 9 | Analyte detection using observation-specific null distributions. A** Partial least squares (PLS) latent-space score plot. Calibration observations are coloured by interferent (sucrose) concentration from grey (low) to brown (high), consistent with Fig. 3C. The blue line denotes the estimated one-dimensional null-space; arrows indicate increasing analyte and interferent levels. Open orange circles and filled green stars show the mean scores of on-grid and off-grid blank test samples, respectively. Plus and cross markers show the mean score of each analyte-containing test sample (samples 5 and 6) and its projected hypothetical blank score, respectively; dotted segments connect each pair. **B** Gaussian-process (GP)-derived (black) and ordinary least squares (OLS)-derived (light blue) null distributions evaluated at the projected mean scores of samples 5 and 6. Histograms show PLS predictions for the 75 observations of the corresponding blank test samples. Coloured vertical lines and arrowheads indicate the predictions for one selected observation from each analyte-containing test sample; black and light-blue dotted vertical lines denote the respective 95% detection limits. LV, latent variable; GP, Gaussian process; OLS, ordinary least squares.

The null-space is established from the 30 blank observations in the calibration set. Since the latent space is two-dimensional and the analyte signal is absent in blank samples, blank variation should be constrained to a one-dimensional null-space (Methods, Analyte detection). This is supported by the mean latent scores of nine independent blank test samples, each consisting of 75 observations, which align closely with the estimated null-space (Fig. 9A).

We next evaluate analyte detection using two test observations: one from sample 5 (0.004 w/w analyte, 0.22 w/w interferent; Supplementary Fig. 4A) and one from sample 6 (0.018 w/w analyte, 0.14 w/w interferent; Supplementary Fig. 4A). Fig. 9A shows their LV scores and projections onto the null-space, and Fig. 9B shows the corresponding null distributions and detection limits ($\alpha = 0.05$). Empirical validation confirms these hypothetical null distributions, and one-sided z-tests successfully detect the analyte in both observations ($p = 0.025$ for sample 5; $p < 0.001$ for sample 6; see Methods, Statistical analyses).

To illustrate how traditional OLS-based variance estimation affects analyte detection when systematic prediction errors are present but ignored, we generate corresponding null distributions under the assumption of zero expected prediction error, assigning all uncertainty to the variance term (Eq. 29, Methods, Statistical analyses). Under this assumption, systematic error is absorbed into the variance term, inflating the variance and raising the detection limit. As Fig. 9B shows, this substantially impairs detection sensitivity at low analyte levels (sample 5).

We then evaluate analyte detection across the 75 observations of samples 5 and 6 and the 75 observations of their corresponding blank samples (Fig. 2; Supplementary Fig. 4A). Type I error analysis (false positive; SI, Supplementary Note 1, Type I and Type II error rates) showed that the OLS uncertainty model yielded Type I $= 0$, whereas the GP uncertainty model yielded Type I $= 0.067$ for both samples 5 and 6 (Table 1). Although a lower Type I error may appear advantageous at first glance, the nominal expectation is given by the significance level ($\alpha = 0.05$), so the zero Type I error under OLS indicates that the null distributions are overly inflated. This interpretation is consistent with Fig. 9B.

**Table 1 | Type I and Type II error rates** for analyte detection in two analyte-containing test samples (0.004 w/w and 0.018 w/w fructose) and their corresponding blank samples, each evaluated over 75 observations. Gaussian process (GP)-based uncertainty incorporates both expected prediction error and variance, whereas ordinary least squares (OLS)-based uncertainty incorporates variance only.

| | **Gaussian process-derived uncertainty (bias and variance)** | | **Ordinary least squares-derived uncertainty (variance only)** | |
|---|---|---|---|---|
| | Type I error | Type II error | Type I error | Type II error |
| Sample 5 (low analyte level) | 0.067 | 0.24 | 0 | 1 |
| Sample 6 (high analyte level) | 0.067 | 0 | 0 | 0 |

Type II error analysis (false negative; SI, Supplementary Note 1, Type I and Type II error rates) showed no false negative observations for sample 6 (high analyte level) under both uncertainty models (Type II $= 0$; Table 1). However, for sample 5 (low analyte level), the OLS uncertainty model failed to detect the analyte in any of the 75 observations (Type II $= 1$; Table 1), whereas the GP uncertainty model correctly detected the analyte in 57 out of the 75 observations (Type II $= 0.24$; Table 1). These results show that the OLS-based null distribution is overly conservative, whereas the GP-based uncertainty model better balances false-positive control and detection sensitivity.

### Generalisation to a quantitative structure-activity relationship dataset

To test whether the framework generalises beyond spectroscopic data, we applied it to a quantitative structure–activity relationship (QSAR) dataset of aqueous solubility predictions from the AqSolDB database;[34,35] the full analysis is presented in SI, Supplementary Note 2. Unlike Raman spectra, QSAR descriptors are heterogeneous calculated molecular properties rather than measured intensities, the latent space is higher-dimensional, and replicate measurements are unavailable. After data splitting and outlier screening, a six-LV PLS model was calibrated on 2,345 compounds, the GP uncertainty framework was trained on a separate validation set of 2,181 compounds, and performance was evaluated on an independent test set of 4,397 compounds.

GP-based error correction reduced the test-set MSE by 22.7 %, from 1.94 to 1.50. A two-sided paired t-test comparing compound-level squared prediction errors before and after correction yielded $t(4396) = 11.14$, $p < 0.001$, and a two-sided Wilcoxon signed-rank test yielded $W > 6 \times 10^6, p < 0.001$. Because the paired squared-error differences were right-skewed and asymmetric (data not shown), these tests are interpreted as exploratory (Supplementary Fig. 8). Supplementary Fig. 9 further shows that the GP-derived uncertainty estimates provided a more informative overall probabilistic description than the corresponding OLS-based estimates: the GP intervals tracked empirical coverage more closely across much of the nominal coverage range, and the GP model assigned observation-specific total prediction-error variances, whereas the OLS approach produced nearly constant variances across compounds. Consistent with this, the GP achieved a lower mean continuous ranked probability score (CRPS)[36] than OLS (0.62 vs 0.76), indicating better overall probabilistic performance. Notably, the GP model achieved this improvement while also producing lower mean and median total variance estimates than OLS (Supplementary Table 2), indicating that the better uncertainty description did not arise from uniformly wider prediction intervals.

These results demonstrate the applicability of the GP frameworks to a structurally different regression problem with a higher-dimensional latent space, non-spectroscopic descriptors, and no replicate measurements.

## Discussion

Latent-variable regression models such as PLS underpin many analytical workflows, yet observation-specific uncertainty estimation is often ignored in practice. Here we show that GP layers trained on PLS latent scores and their associated prediction errors can recover both the local expected error and error variance of individual predictions. Using a two-constituent system, we demonstrate four main advances.

First, observation-specific systematic errors can be substantial in LV-based regression, but can be reduced when their local structure is learned. In our Raman example, prediction errors changed sign across the latent space because spectral overlap caused the two-LV PLS model to trade off analyte and interferent contributions differently across the calibration range; residuals were smallest near the calibration mean and increased towards the extremes of the latent space. Because the error GP learns this local structure from the latent scores, it can reduce systematic prediction errors in an observation-specific manner. This reduced MSE by approximately 59% in both Raman test sets, indicating that a substantial component of the prediction error was systematic and could be reduced through observation-specific correction.

Second, our GP-derived variance estimates provide sensible, observation-specific prediction intervals without over-inflation. Total prediction-error variance varies systematically across the latent space, increasing from the lower-left to the upper-right region of the score plot, corresponding to increasing analyte and interferent levels (Fig. 4B). Decomposing the total prediction-error variance into epistemic and aleatoric components shows that this pattern is driven primarily by the aleatoric term, whereas epistemic variance remains small and increases only modestly toward the boundaries of the latent space (Fig. 5). Thus, the dominant source of uncertainty in this system is composition-dependent heteroscedasticity rather than lack of local model support. Consistent with this, Fig. 7B shows broad agreement between GP-derived and empirically estimated aleatoric variances across both on-the-grid and off-the-grid test samples. One plausible explanation for the apparent increase in aleatoric variance toward higher analyte and interferent levels is that absolute uncertainties associated with sample preparation and measurement scale with concentration.

Third, bias-aware uncertainty enables more reliable detection limits in multiplex settings. By projecting each observation onto the blank subspace and constructing a tailored null distribution that incorporates both local expected prediction error and local variance, the detection threshold becomes observation-specific rather than globally fixed. In our Raman example, this was particularly important for the low-analyte sample, where the GP-based uncertainty model yielded low Type II error whereas the variance-only OLS approach failed completely to detect the analyte (Table 1). By contrast, both approaches performed well for the higher-analyte sample, where the signal was sufficiently separated from the corresponding blank distribution. These results show that even small systematic errors can substantially reduce sensitivity near the decision boundary, and that explicitly modelling both expected prediction error and variance can materially improve analytical performance.

The same principle could be extended to partial least squares regression with multiple response variables (PLS2), which enables several analytes to be calibrated simultaneously.[27] Because the responses share the same latent scores, separate error and variance GP models could be fitted to the prediction errors for each analyte. Analyte-specific detection limits would then require calibration samples in which the target analyte is absent while the remaining analytes and interferents span the intended application domain, thereby defining a separate null subspace for each analyte. For simultaneous decisions concerning multiple analytes, correlations between their prediction errors would also need to be considered, for example using a multi-output GP.

Fourth, the latent-space view provides mechanistic insight into the source of prediction errors. Decomposing the prediction into LV1 and LV2 contributions (Supplementary Fig. 5) clarifies how interferent levels perturb analyte estimates and why the resulting error surface becomes saddle-shaped. Such structure is difficult to extract from bootstrap- or jackknife-based uncertainty estimates, but follows naturally from modelling prediction bias and variance directly in the latent space. These diagnostics may help guide domain-specific countermeasures, such as improved experimental design, targeted preprocessing, or model redesign.

Finally, the framework also generalises beyond the spectroscopic setting. In the QSAR aqueous-solubility example, the inputs were heterogeneous molecular descriptors rather than measured spectral intensities; the latent space was higher-dimensional; and replicate measurements were unavailable. Yet, despite these differences, the error GP captured systematic prediction error and improved predictive accuracy after correction, and the variance GP produced observation-specific uncertainty estimates that outperformed the corresponding OLS-based uncertainties in overall probabilistic performance (Supplementary Note 2). Notably, this improvement was achieved even though the GP model yielded lower mean and median total variance estimates than OLS, indicating that the improved uncertainty description did not arise from uniformly wider prediction intervals, but from better-distributed uncertainty across observations. The smaller gain from GP-derived error correction in the QSAR case may reflect the PLS bias–variance trade-off. Indeed, the QSAR model used more latent variables, hence systematic error was likely already reduced in the base PLS model, leaving less residual bias for the GP layer to correct. Together, these results support the broader applicability of the framework beyond spectroscopy.

Importantly, the framework does not require the underlying data to be intrinsically low-rank. Additional independently varying interferents may increase the effective rank of the data and the number of latent variables required, but the Gaussian-process models operate on the latent representation provided by the LV model. The QSAR example demonstrates that the framework can operate with a higher-dimensional latent representation: its inputs were heterogeneous molecular descriptors and the PLS model required six LVs. The framework may also accommodate multiple interferents, provided that the calibration and validation data adequately cover the relevant combinations of analyte and interferent levels and that their effects are represented in the retained latent scores. In higher-dimensional latent spaces, the resulting error structures cannot generally be visualised as a single surface. They could instead be examined using conditional two-dimensional slices or one-dimensional profiles, with the remaining scores fixed at representative values. When interferent concentrations are known, a series of plots for selected combinations of interferent levels could provide a complementary interpretation.

Prediction intervals are expected to be most reliable within, or close to, the latent-score region supported by the training data. Near the boundary of that region, the estimates may still be informative, but should be interpreted with caution because epistemic variance increases as local support weakens. For samples clearly outside the calibration domain, the estimates become extrapolative: with the RBF kernel used here, expected prediction error and aleatoric variance tend to revert toward baseline levels learned from the training data, while epistemic variance increases toward a limit value. Extrapolative behaviour is therefore kernel-dependent, and alternative kernels could be used if different behaviour outside the calibration domain is desired.[28]

Because the GP layer is trained on the residual structure of the underlying PLS model, its validity depends on the validity of that PLS model. If new interferents or other sources of variation not represented in the calibration data are introduced, the latent-space mapping itself may no longer be reliable, and the GP-derived expected prediction error and variance estimates should not be assumed to be valid automatically. In practice, such cases could be screened using conventional applicability-domain diagnostics, such as score-space distance measures related to Hotelling's $T^2$ and $Q$ residuals from PCA or PLS,[37] although we did not implement such criteria in the Raman analysis.

The main practical limitation of our approach is computational: standard GP training scales as $\mathcal{O}(N^3)$ with the number of training observations,[28] and the cost may increase further with latent-space dimensionality. For problems with many observations or higher-dimensional latent spaces, sparse GP approximations and related scalable variants may therefore be needed.[38,39] Hyperparameter optimisation is also important, because poor optimisation may either oversmooth latent-space trends or fit spurious local structure; here we used marginal-likelihood optimisation with multiple restarts to reduce the risk of poor local solutions.

No universal minimum sample size can be prescribed, because the required number of observations depends on the dimensionality and geometric complexity of the latent space, as well as on how rapidly the expected prediction error and variance vary across it. In practice, the training data should provide sufficient coverage of the latent-score regions over which uncertainty is to be learned. In the special case of homoscedastic data, the aleatoric variance may be approximated as constant, in which case the variance term could be simplified to a single noise hyperparameter in the error GP rather than a separate variance GP.[28]

Maximum-likelihood principal component regression (MLPCR) explicitly accounts for predictor-block measurement error during model fitting.[40] This may reduce prediction-error structure caused by measurement error in the predictors and its propagation into the latent scores. However, MLPCR still relies on a reduced latent representation and would therefore not necessarily eliminate score-dependent prediction error. Score-dependent expected error or variance may remain because dimensionality reduction can discard response-relevant information, or because of model mismatch, nonlinear relationships, unmodelled interferents, heteroscedastic noise, or errors in the reference response values. The proposed GP layer could therefore complement MLPCR by modelling any remaining expected prediction error and residual variance across the latent space.

More broadly, the framework provides a practical route for extending LV regression from point prediction to observation-specific probabilistic prediction while retaining the interpretability of the latent-space representation. Its main practical advantages are that it is post-hoc, requires no change to the underlying regression algorithm, yields a full predictive distribution rather than only an asymptotic variance, and can in principle be trained on scores from other LV regression models, such as principal component regression or nonlinear PLS.[41,42]

## Methods

### Partial least squares regression

PLS regression was used to project the explanatory variables, $\mathbf{X} \in \mathbb{R}^{N \times M}$ and the response variable, $\mathbf{y} \in \mathbb{R}^{N}$ onto a shared, lower-dimensional latent space.[10,27] Using the fitted PLS model, predictions are obtained from $\mathbf{T}$ as

$$\hat{\mathbf{y}} = \mathbf{T}\mathbf{q} \tag{3}$$

where $\hat{\mathbf{y}} \in \mathbb{R}^{N}$ are analyte predictions; and $\mathbf{q} \in \mathbb{R}^{A}$ is the latent space regression coefficients.[27] A fuller description of PLS and its application to the Raman model system is provided in the SI (Supplementary Note 1, Partial least squares regression).

The resulting $\mathbf{T}$ served as input to the GP pipeline, while the corresponding prediction errors, $\mathbf{e} = \mathbf{y} - \hat{\mathbf{y}}$; $\mathbf{e} \in \mathbb{R}^{N}$, were used as outputs.

### Prediction-error decomposition

The prediction error for observation $i$, denoted $e_i$, is defined in Eq. 2. We decompose this error into an expected prediction error, an epistemic variance, and an aleatoric variance.

Randomness in $y_i$ may arise from inaccuracies in analytical instruments, sample preparation errors, or environmental factors.[15] Formally,

$$y_i = \mathbb{E}[y_i] + \varepsilon_i, \qquad \varepsilon_i \sim \mathcal{N}(0, \sigma_{\varepsilon_i}^2) \tag{4}$$

where $\mathbb{E}[y_i]$ is the expected $y_i$-value; and $\varepsilon_i$ is a zero-mean random variable with variance $\sigma_{\varepsilon_i}^2$. Repeated measurements of sample $i$ therefore yield a distribution of $y_i$-values centred on $\mathbb{E}[y_i]$, with variance $\sigma_{\varepsilon_i}^2$. Hence, $\sigma_{\varepsilon,i}^2$ reflects aleatoric variability in $y_i$.

Randomness in $\hat{y}_i$ may arise from measurement noise in the explanatory-variable vector $\mathbf{x}_i \in \mathbb{R}^M$ for observation $i$ and from uncertainty in the local regression mapping. Formally,

$$\hat{y}_i = \mathbb{E}[\hat{y}_i] + \eta_i, \qquad \eta_i \sim \mathcal{N}(0, \sigma_{\eta_i}^2) \tag{5}$$

where $\mathbb{E}[\hat{y}_i]$ is the expected $\hat{y}_i$-values; and $\eta_i$ is a zero-mean random variable with variance $\sigma_{\eta_i}^2$. We further decompose $\sigma_{\eta_i}^2$ into aleatoric, $\sigma_{\eta_i,\text{ale}}^2$, and epistemic, $\sigma_{\text{epi}}^2$, components,

$$\sigma_{\eta_i}^2 = \sigma_{\eta_i,\text{ale}}^2 + \sigma_{\text{epi}}^2 \tag{6}$$

where $\sigma_{\eta_i,\text{ale}}^2$ reflects measurement noise in $\mathbf{x}_i$ that propagates through the regression mapping to $\hat{y}_i$; and $\sigma_{\text{epi}}^2$ reflects uncertainty in the estimate of $\mathbb{E}[\hat{y}_i]$ and typically increases in sparsely supported latent-space regions.

The systematic deviation between $\mathbb{E}[y_i]$ and $\mathbb{E}[\hat{y}_i]$ defines the expected prediction error, $\mathbb{E}[e_i]$,

$$\mathbb{E}[e_i] = \mathbb{E}[y_i] - \mathbb{E}[\hat{y}_i] \tag{7}$$

which equals the negative of the bias.

The error distribution of observation $i$ is obtained by substituting Eq. 4, 5, and 7 into Eq. 2,

$$e_i = y_i - \hat{y}_i = (\mathbb{E}[y_i] - \mathbb{E}[\hat{y}_i]) + (\varepsilon_i - \eta_i) = \mathbb{E}[e_i] + (\varepsilon_i - \eta_i) \tag{8}$$

which shows that the error distribution is centred at $\mathbb{E}[e_i]$ and the random dispersion arises from $\varepsilon_i$ and $\eta_i$. Assuming that $\varepsilon_i$ and $\eta_i$ are uncorrelated conditional on the latent-score location, the error variance, $\sigma_{e_i}^2 = \sigma_{\varepsilon_i}^2 + \sigma_{\eta_i}^2$, is then expressed in terms of its aleatoric and epistemic parts

$$\sigma_{e_i}^2 = \sigma_{\text{ale}_i}^2 + \sigma_{\text{epi}_i}^2 \tag{9}$$

where $\sigma_{\text{ale}_i}^2 = \sigma_{\varepsilon_i}^2 + \sigma_{\eta_i,\text{ale}}^2$. This decomposition follows the law of total variance and is conditional on $\mathbf{t}_i$.

Error distributions are observation-specific because LV regression introduces a bias–variance trade-off through dimensionality reduction, and because regression-to-the-mean effects depend on LV location. Thus, nearby observations in latent space are expected to share similar error distributions. In the following section, we use Gaussian-process regression to estimate these three quantities, $\mathbb{E}[e(\mathbf{t})]$, $\sigma_{\text{epi}}^2(\mathbf{t})$, and $\sigma_{\text{ale}}^2(\mathbf{t})$, as continuous functions of latent-space location, $\mathbf{t}$.

### Gaussian-process models

GP regression was used to model prediction-error structure over the latent space defined by the PLS scores, $\mathbf{t}$. We fitted two GP regression models to the validation data, both using latent scores as input. The first—the error GP—estimates $\mathbb{E}[e(\mathbf{t})]$ and $\sigma^2_{\text{epi}}(\mathbf{t})$. The second—the variance GP—estimates $\sigma^2_{\text{ale}}(\mathbf{t})$.

For both GP models, covariance between two observations with PLS latent scores $\mathbf{t}_i \in \mathbb{R}^A$ and $\mathbf{t}_j \in \mathbb{R}^A$ was described by an automatic relevance determination (ARD) radial basis function (RBF) kernel,

$$k(\mathbf{t}_i, \mathbf{t}_j) = \sigma_f^2 \exp\left(-\sum_{a=1}^{A}\left(\frac{(t_{i,a}-t_{j,a})^2}{2\ell_a^2}\right)\right) \quad (10)$$

where $\sigma_f^2$ is the signal variance that controls the scale of prediction-error covariance; $\ell_a$ is the length scale for LV $a$, which determines how quickly prediction-error covariances decay with increasing distance along that dimension; and $A$ is the number of latent variables in the PLS model. We denote the error GP kernel as $k_e$ with hyperparameters $\sigma_{e_f}$ and $\{\ell_{e_a}\}_{a=1}^{A}$, and the variance GP kernel as $k_v$ with hyperparameters $\sigma_{v_f}$ and $\{\ell_{v_a}\}_{a=1}^{A}$. For the two GP models, hyperparameters were independently optimised by minimising the negative log marginal likelihood (exact for the error GP; Laplace-approximated for the variance GP).

The error GP was fitted to the $(\mathbf{t}_i, e_i)$ pairs from the validation data, where $e_i$ is the PLS prediction error of observation $i$ (Eq. 2). The posterior mean, which estimates the expected prediction error, $\widehat{\mathbb{E}}[e(\mathbf{t}_i)]$, at latent-space location $\mathbf{t}_i$, is given by

$$\widehat{\mathbb{E}}[e(\mathbf{t}_i)] = \mathbf{k}_{e_i}^{\mathrm{T}}(\mathbf{K}_e + \sigma_n^2\mathbf{I}_N)^{-1}\mathbf{e} \quad (11)$$

where $\mathbf{K}_e + \sigma_n^2\mathbf{I}_N$ corresponds to the covariance matrix of prediction errors across the latent space; $\mathbf{K}_e \in \mathbb{R}^{N\times N}$ is the matrix of pairwise kernel evaluations $k_e(\mathbf{t}_i, \mathbf{t}_j)$ (Eq. 10) of validation observations; $\sigma_n^2$ is the error-GP noise-hyperparameter, optimised jointly with the kernel hyperparameters; $\mathbf{I}_N \in \mathbb{R}^{N\times N}$ is the identity matrix; $\mathbf{e} \in \mathbb{R}^N$ is the vector of prediction errors of the validation data; and $\mathbf{k}_{e_i} \in \mathbb{R}^N$ contains the covariances between each validation observation and observation $i$ (obtained from Eq. 10).

Furthermore, the error GP is used to estimate the epistemic variance, $\hat{\sigma}^2_{\text{epi}}(\mathbf{t}_i)$ (i.e., the uncertainty about $\widehat{\mathbb{E}}[e(\mathbf{t}_i)]$). This is given by[28]

$$\hat{\sigma}^2_{\text{epi}}(\mathbf{t}_i) = k_e(\mathbf{t}_i, \mathbf{t}_i) - \mathbf{k}_{e_i}^{\mathrm{T}}(\mathbf{K}_e + \sigma_n^2\mathbf{I}_N)^{-1}\mathbf{k}_{e_i}. \quad (12)$$

To estimate the aleatoric variance (the variance GP model), we first debiased the prediction errors using the error GP,

$$\tilde{e}_i = e_i - \widehat{\mathbb{E}}[e(\mathbf{t}_i)] \quad (13)$$

where $\tilde{e}_i$ denotes the debiased residual and $\widehat{\mathbb{E}}[e(\mathbf{t}_i)]$ is the expected error evaluated at $\mathbf{t}_i$ using Eq. 11. Hence, $\tilde{e}_i$ reflect stochastic fluctuation about $\widehat{\mathbb{E}}[e(\mathbf{t}_i)]$, which may vary across the latent space under heteroscedasticity.

This second stage follows the input-dependent-noise Gaussian-process framework, in which the log noise variance is assigned a GP prior, as introduced by Goldberg et al.[29] and later approximated efficiently in heteroscedastic GP regression by Kersting et al.[30] Here, we use a two-stage plug-in formulation with a Laplace approximation, following the standard Gaussian-process treatment of non-Gaussian likelihoods, to obtain the posterior mean and variance of the latent log-variance function.[28]

We then modelled the latent log-variance function $g(\mathbf{t}) = \ln(\sigma^2_{\text{ale}}(\mathbf{t}))$ with a GP prior,

$$g(\mathbf{t}) \sim \mathcal{GP}\left(\ln\left(\text{median}(\tilde{\mathbf{e}}^2)\right), k_v(\mathbf{t}, \mathbf{t}')\right) \tag{14}$$

where $\ln\left(\text{median}(\tilde{\mathbf{e}}^2)\right)$ is a constant prior mean set from the validation data, and $k_v$ is the variance GP kernel of the form given in Eq. 10. The debiased residuals are related to $g(\mathbf{t})$ through the observation model

$$\tilde{e}_i | g(\mathbf{t}_i) \sim \mathcal{N}\left(0, \exp\left(g(\mathbf{t}_i)\right)\right). \tag{15}$$

Because this likelihood is non-Gaussian in the latent log-variance vector $\mathbf{g} = [g(\mathbf{t}_1), \ldots, g(\mathbf{t}_N)]^{\text{T}} \in \mathbb{R}^N$, the posterior over $\mathbf{g}$ has no closed-form solution. We therefore used a Laplace approximation, in which the posterior is approximated as Gaussian around its maximum a posteriori (MAP) estimate, $\hat{\mathbf{g}}$.

The posterior mean of $\mathbf{g}$ at latent-space location $\mathbf{t}_i$ is

$$\mathbb{E}[g(\mathbf{t}_i)|\tilde{e}_i] = \ln\left(\text{median}(\tilde{\mathbf{e}}^2)\right) + \mathbf{k}_{v_i}^{\text{T}} \mathbf{K}_v^{-1}\left(\hat{\mathbf{g}} - \ln\left(\text{median}(\tilde{\mathbf{e}}^2)\right)\mathbf{1}_N\right) \tag{16}$$

where $\mathbf{k}_{v_i} \in \mathbb{R}^N$ is the vector of covariances between the log-variance at each validation location and the log-variance at location $\mathbf{t}_i$, evaluated using the variance GP kernel $k_v$ (Eq. 10); $\mathbf{K}_v \in \mathbb{R}^{N \times N}$ is the matrix of pairwise kernel evaluations among validation locations; $\hat{\mathbf{g}} \in \mathbb{R}^N$ is the MAP estimate of $\mathbf{g}$; and $\mathbf{1}_N \in \mathbb{R}^N$ is a vector of ones.

The posterior variance of $g(\mathbf{t}_i)$ under Laplace approximation is given by

$$\sigma^2_{g(\mathbf{t}_i)} = k_v(\mathbf{t}_i, \mathbf{t}_i) - \mathbf{s}_i^{\text{T}} \mathbf{B}^{-1} \mathbf{s}_i \tag{17}$$

where $k_v(\mathbf{t}_i, \mathbf{t}_i)$ is the prior variance of $\mathbf{g}$ at location $\mathbf{t}_i$, and the second term quantifies the variance reduction due to the observed data. Here $\mathbf{s}_i = \mathbf{\Lambda}^{1/2}\mathbf{k}_{v_i}$; $\mathbf{B} = \mathbf{I}_N + \mathbf{\Lambda}^{1/2}\mathbf{K}_v\mathbf{\Lambda}^{1/2}$, where $\mathbf{\Lambda} = \text{diag}(\Lambda_{ii}) \in \mathbb{R}^{N \times N}$ is a diagonal matrix whose entries $\Lambda_{ii} = \frac{1}{2}\tilde{e}_i^2 \exp(-\hat{g}_i)$ are the negative second derivatives of the log-likelihood at the MAP. The aleatoric variance is then estimated via the log-normal mean

$$\hat{\sigma}^2_{\text{ale}}(\mathbf{t}_i) = \exp\left(\mathbb{E}[g(\mathbf{t}_i)] + \frac{1}{2}\sigma^2_{g(\mathbf{t}_i)}\right). \tag{18}$$

Together, the error GP (Eqs. 10–12) and variance GP (Eqs. 16–18) provide estimates of all three components of the prediction-error distribution (Eqs. 7–9): the expected prediction error, the epistemic variance, and the aleatoric variance.

## Model testing

The GP models were evaluated using the test set (Fig. 2). The Raman spectrum of a given test observation was first preprocessed (SI, Supplementary Note 1, Spectral preprocessing) and then mean-centred using the spectral mean of the preprocessed calibration set. The mean-centred spectrum, $\mathbf{x}_* \in \mathbb{R}^M$, was subsequently projected onto the calibrated PLS model using Eq. 1 to obtain its LV scores, $\mathbf{t}_* \in \mathbb{R}^A$. These scores were then passed to the error GP to obtain the expected prediction error (Eq. 11) and epistemic variance (Eq. 12), and to the variance GP to obtain the aleatoric variance (Eqs. 16–18).

For the comparative evaluation of GP-derived and empirically determined prediction-error distributions, the LV scores were averaged across the 75 test-set observations for each test sample. These mean scores were then input into the GP models to obtain GP-derived prediction-error distributions, which were directly compared to empirically estimated distributions derived from the corresponding test-set observations (Fig. 6, Fig. 7, Supplementary Fig. 4B).

**Gaussian-process-corrected predictions**

When $\widehat{\mathbb{E}}[e_*]$ is obtained from the error GP (Eq. 11), the corrected prediction, $\hat{y}_*^{\text{corr}}$, is obtained as

$$\hat{y}_*^{\text{corr}} = \hat{y}_* + \widehat{\mathbb{E}}[e_*] \quad (19)$$

where $\hat{y}_* = \mathbf{t}_*^{\mathrm{T}}\mathbf{q}$ is the PLS analyte prediction of the new observation.

**Prediction interval**

The prediction interval for $\hat{y}_*^{\text{corr}}$, $\text{CI}_*$ is then obtained as

$$\text{CI}_* = \hat{y}_*^{\text{corr}} \pm z_{1-\alpha/2}\hat{\sigma}_{e_*} \quad (20)$$

where $z_{1-\alpha/2}$ is the critical value of the standard normal distribution for the desired confidence level, $1-\alpha$; and $\hat{\sigma}_{e_*} = \sqrt{\sigma_{\text{epi}_*}^2 + \sigma_{\text{ale}_*}^2}$ is the estimated total prediction-error standard deviation (Eq. 9), with the epistemic component ($\sigma_{\text{epi}_*}^2$) from Eq. 12 and the aleatoric component ($\sigma_{\text{ale}_*}^2$) from Eq. 18, evaluated at latent space location $\mathbf{t}_*$.

**Analyte detection**

To robustly detect analyte presence, we first establish a null-space that captures the variation inherent to blank samples. Consider $N_{\text{Blank}}$ blank observations (without the analyte), then we define the null-space by the scores of blank samples, $\mathbf{T}_{\text{Blank}} \in \mathbb{R}^{N_{\text{Blank}} \times A}$, whose distribution depends on interferent levels. For any new sample, we can then project its latent scores onto the null-space to derive its corresponding hypothetical blank scores. Note that in the PLS latent space, the analyte is assumed to be captured by a single dimension; hence blank samples—free from analyte signal—exhibit variation in the remaining $A-1$ dimensions of the full $A$-dimensional latent space.[32]

The basis of the null-space, $\mathbf{Q}_{\text{Blank}} \in \mathbb{R}^{A \times (A-1)}$, can be defined using singular value decomposition on the column centred $\mathbf{T}_{\text{Blank}}$,

$$\mathbf{T}_{\text{Blank}} - \mathbf{1}_{N_{\text{Blank}}}\bar{\mathbf{t}}_{\text{Blank}}^{\mathrm{T}} = \mathbf{USV}^{\mathrm{T}} \quad (21)$$

where $\mathbf{1}_{N_{\text{Blank}}} \in \mathbb{R}^{N_{\text{Blank}}}$ denotes a column vector of ones; $\bar{\mathbf{t}}_{\text{Blank}} \in \mathbb{R}^{A}$ is a vector containing column means of $\mathbf{T}_{\text{Blank}}$; $\mathbf{U} \in \mathbb{R}^{N_{\text{Blank}} \times N_{\text{Blank}}}$ is an orthonormal matrix whose columns are the left singular vectors; $\mathbf{S} \in \mathbb{R}^{N_{\text{Blank}} \times A}$ is a diagonal matrix containing the singular values; $\mathbf{V} \in$

$\mathbb{R}^{A\times A}$ is an orthonormal matrix whose columns are the right singular vectors, representing the principal directions in $\mathbf{T}_{\text{Blank}}$. Now, $\mathbf{Q}_{\text{Blank}}$ is defined as the first $A-1$ columns of $\mathbf{V}$,

$$\mathbf{Q}_{\text{Blank}} = [\mathbf{v}_1, \mathbf{v}_2, \cdots, \mathbf{v}_{A-1}]. \quad (22)$$

To project a new observation onto the null-space, one must choose a projection direction. Allegrini and Olivieri project calibration samples perpendicularly onto the zero-analyte hyperplane defined in score space by the PLS regression coefficient vector.[32] However, an orthogonal projection only preserves a sample's interferent-related coordinates when the analyte direction in score space is perpendicular to the null-space. When analyte and interferent signals overlap spectrally, this condition is generally not met, and the projected point corresponds to a hypothetical blank with a different interferent background than the original sample. We therefore perform an oblique projection along the estimated analyte direction, which preserves the sample's interferent coordinates and traces it back to its physically correct hypothetical blank conditions. To estimate this direction, we fit a linear model relating the calibration scores to analyte and interferent levels

$$\mathbf{t}_i \approx \boldsymbol{\beta}_0 + \boldsymbol{\beta}_z \cdot z_i + \boldsymbol{\beta}_y \cdot y_i \quad (23)$$

where $\boldsymbol{\beta}_0 \in \mathbb{R}^A$ is the intercept; $\boldsymbol{\beta}_z \in \mathbb{R}^A$ captures the effect of interferent concentration $z_i$; and $\boldsymbol{\beta}_y \in \mathbb{R}^A$ captures the effect of analyte concentration $y_i$. The analyte direction is then defined as

$$\mathbf{u}_y = \frac{\boldsymbol{\beta}_y}{|\boldsymbol{\beta}_y|}. \quad (24)$$

To assess analyte detection for a new sample with latent scores $\mathbf{t}_*$, we perform an oblique projection along the analyte direction onto the null-space. Specifically, we decompose the centred scores into null-space and analyte-direction components

$$\mathbf{t}_* - \bar{\mathbf{t}}_{\text{Blank}} = \mathbf{Q}_{\text{Blank}}\boldsymbol{\xi} + \mathbf{u}_y\gamma \quad (25)$$

where $\boldsymbol{\xi} \in \mathbb{R}^{A-1}$ denotes the coefficients in the null-space basis $\mathbf{Q}_{\text{Blank}}$; and $\gamma$ is the coefficient (scalar) along the analyte direction. Solving this linear system for $\boldsymbol{\xi}$ yields the hypothetical blank scores

$$\mathbf{t}_{*,\text{null}} = \bar{\mathbf{t}}_{\text{Blank}} + \mathbf{Q}_{\text{Blank}}\boldsymbol{\xi}. \quad (26)$$

The resulting $\mathbf{t}_{*,\text{null}} \in \mathbb{R}^A$ traces the sample back along the analyte direction to the null-space, representing the hypothetical blank conditions for the new sample.

An earlier approach used the leverage of hypothetical blank sample to estimate PLS-based prediction variance, from which a confidence interval for the hypothetical blank sample was calculated.[32] However, as Allegrini and Olivieri point out, this method assumes zero prediction bias and assigns all uncertainty to variance alone.[32] When systematic bias is present, this variance estimate becomes inflated, producing overly conservative confidence intervals that can obscure low-level analyte detection.

Within our framework, we instead propose using the projected latent scores, $\mathbf{t}_{*,\text{null}}$, as input to the GP model to obtain the expected prediction error, $\hat{\mathbb{E}}[e_{*,\text{null}}]$ and the prediction-error variance, $\hat{\sigma}^2_{e_{*,\text{null}}}$, for the hypothetical blank sample. Since the expected response value for the hypothetical blank sample is $\mathbb{E}[y_{*,\text{null}}] = 0$, the expected estimated response for that sample follows from Eq. 7 (Methods, Prediction-error decomposition) as, $\mathbb{E}[\hat{y}_{*,\text{null}}] = -\hat{\mathbb{E}}[e_{*,\text{null}}]$.

Therefore, we can define $\mathrm{DL}_*$, the analyte detection limit for $\hat{y}_*$ taking into account the prediction bias, as

$$\mathrm{DL}_* = -\hat{\mathbb{E}}\left[e_{*,\mathrm{null}}\right] + z_{1-\alpha}\hat{\sigma}_{e_{*,\mathrm{null}}} \tag{27}$$

where $z$ is the critical value of the normal distribution and $\alpha$ is the significance level.

## Statistical analyses

The effect of GP-derived correction was evaluated separately for the on-the-grid and off-the-grid test sets, comprising 25 and 12 mixtures, respectively, with 75 observations per mixture. GP correction was applied to each observation using its own latent scores (Eq. 19). Squared prediction errors before and after correction were then averaged within each mixture. Each mixture therefore contributed one paired MSE comparison:

$$d_j = \frac{1}{n_j}\sum_{i=1}^{n_j}\left((y_{ij} - \hat{y}_{ij})^2 - (y_{ij} - \hat{y}_{ij}^{\mathrm{corr}})^2\right). \tag{28}$$

Here, $j$ indexes mixtures, $i$ indexes observations within a mixture, and $n_j$ is the number of test observations for that mixture ($n_j = 75$ for every mixture in this study). Positive $d_j$ indicates a reduction in MSE after correction.

Two-sided paired t-tests assessed whether the mean mixture-level MSE difference was zero, yielding $t(24) = 3.19, p = 0.0040$ for the on-the-grid test set and $t(11) = 2.33, p = 0.0399$ for the off-the-grid test set. Two-sided Wilcoxon signed-rank tests yielded $W = 307$, $p = 0.0001$ and $W = 72$, $p = 0.0068$, respectively, where $W$ is the sum of the ranks of positive differences. Both tests assume independent mixture-level differences. Exact inference from the paired t-test assumes normally distributed differences, whereas the Wilcoxon signed-rank test assumes a continuous distribution symmetric about its median. Inspection of the mixture-level MSE differences indicated right-skewness and asymmetry (data not shown). The resulting $p$-values are therefore interpreted as exploratory and with caution, particularly for the off-the-grid test set, which comprised only 12 mixtures.

Analyte detection was performed using a one-sided z-test ($H_0\!: y_i = 0$ vs. $H_1\!: y_i > 0$) comparing each of the 75 observations of sample 5 and sample 6 (Supplementary Fig. 4A) against their estimated null distributions with the detection limit set at $\alpha = 0.05$. Two alternative null-distribution assumptions were compared: The GP-based null distribution, $\mathcal{N}(-\hat{\mathbb{E}}[e_{i,\mathrm{null}}], \hat{\sigma}_{i,\mathrm{null}}^2)$, accounting for the expected prediction error, where $\hat{\mathbb{E}}[e_{i,\mathrm{null}}]$ and $\hat{\sigma}_{i,\mathrm{null}}^2$ are returned from inputting the observation-of-interest's hypothetical blank scores (Eq. 26 in Methods, Analyte detection) into the GP models (Eqs. 11–12 and Eqs. 16–18 in Methods, Gaussian-process models); and the OLS-rooted null-distribution, assuming unbiased predictions, $\mathcal{N}(0, {\hat{\sigma}_{i,\mathrm{null}}^2}^{\mathrm{OLS}})$,

$${\hat{\sigma}_{i,\mathrm{null}}^2}^{\mathrm{OLS}} = \left(1 + \frac{1}{N} + \mathbf{t}_{i,\mathrm{null}}^{\mathrm{T}}(\mathbf{T}^{\mathrm{T}}\mathbf{T})^{-1}\mathbf{t}_{i,\mathrm{null}}\right)\mathrm{MSE}^{\mathrm{loocv}} \tag{29}$$

where $N = 150$ is the number of calibration observations; $\mathbf{T}$ contains their latent scores; $\mathbf{t}_{i,\mathrm{null}}$ is the projected latent scores of observation $i$ under the null hypothesis (Eq. 26, Methods, Analyte detection); and $\mathrm{MSE}^{\mathrm{loocv}}$ is the leave-one-out CV MSE.[3,32]

## Data Availability

All raw Raman spectra (.csv) and corresponding reference values are available under a CC BY 4.0 licence.[43] The processed calibration, validation and test Raman datasets, including their reference values and split assignments, are included in the repository described in the Code Availability statement and are likewise available under a CC BY 4.0 licence.[44] The third-party AqSolDB dataset used for the QSAR analysis is not redistributed in that repository and is publicly available from Harvard Dataverse under a CC0 licence.[35]

## Code Availability

MATLAB R2022a scripts and functions for PLS fitting, heteroscedastic GP training, and figure generation are available at https://github.com/ceeskildsen/Supporting-Software-for-Observation-Specific-Uncertainty-Modeling-for-Latent-Variable-Regression. A versioned snapshot of the complete supporting software, Raman data and derived analysis outputs is archived on Zenodo.[44] The heteroscedastic GP pipeline is additionally archived separately on Zenodo.[45] All code is released under the MIT licence. Running the scripts requires MATLAB R2022a or later, with the Statistics and Machine Learning Toolbox and the Optimization Toolbox installed.

## Acknowledgements

Dr Álvaro Fernandez-Galiana is acknowledged for assisting during Raman data collection. Ms Elena Khachirova is acknowledged for proofreading the manuscript. Dr Akemi Nogiwa is acknowledged for proofreading the manuscript and checking the data-management documentation.

## Funding Statement

The Engineering and Physical Sciences Research Council (EPSRC) under the UK Research and Innovation (UKRI) is acknowledged for funding C.E.E. through the project Trustworthy Decision Limits for Multiplexed Diagnostics (Trust-MDx; EP/Y023498/1), which was granted via the Horizon Europe Marie Skłodowska-Curie Actions individual fellowship programme and funded by the UKRI Horizon Europe guarantee scheme. G.P. is supported by the UKRI Centre for Doctoral Training in AI for Healthcare (EP/S023283/1). K.A.G declares no relevant funding. M.B. acknowledges support from the Engineering and Physical Sciences Research Council (EPSRC; EP/N014529/1, funding the EPSRC Centre for Mathematics of Precision Healthcare at Imperial, and EP/T027258/1). M.M.S. acknowledges support from the Department for Science, Innovation and Technology (DSIT) and the Royal Academy of Engineering under the Chair in Emerging Technologies programme (CiET2021\94), the University of Oxford Strategic Research Fund and the Research Council of Norway (262613).

## Author Contributions Statement

C.E.E. conceived and designed the research, oversaw sample preparation and data collection, and was responsible for data analysis, interpretation of results, and manuscript writing, ensuring the overall integrity of the work. G.P. contributed to data analysis and interpretation of results. K.A.G. contributed to sample preparation and data collection. M.B. contributed to manuscript writing and the interpretation of results. M.M.S. secured and coordinated access to core laboratory facilities and instrumentation used for data acquisition and supervised the project. All authors reviewed the manuscript.

## Competing Interests Statement

M.M.S. invested in, consults for (or was on scientific advisory boards or boards of directors), conducts sponsored research funded by companies related to the biomaterials and Raman spectroscopy field (SPARTA Biodiscovery Ltd.), and is inventor in patents 2314282.1/GB/PRV and EP3811052A1 describing Raman techniques (unrelated to the work presented here). The rest of the authors declare no competing interests.

## Additional Information

The online version contains supplementary materials.

Correspondence and requests for materials should be addressed to Carl Emil Eskildsen or Molly M. Stevens.

# SUPPLEMENTARY INFORMATION
# FOR

## Observation-Specific Uncertainty Modeling for Latent-Variable Regression in High-Dimensional Analytical Data

Carl Emil Eskildsen[1,2,3‡], Georgios Papadopoulos[2,3,4], Kevin A. Giraldo[2,3], Mauricio Barahona[5], Molly M. Stevens[1,2,3,6,7,8‡]

[1]Department of Bioengineering, Imperial College London, London SW7 2AZ, United Kingdom
[2]Department of Physiology, Anatomy and Genetics, University of Oxford, Oxford OX1 3PT, United Kingdom
[3]Kavli Institute for Nanoscience Discovery, University of Oxford, Oxford OX1 3QU, United Kingdom
[4]Department of Computing, Imperial College London, London SW7 2AZ, United Kingdom
[5]Department of Mathematics, Imperial College London, London SW7 2AZ, United Kingdom
[6]Department of Materials, Imperial College London, London SW7 2AZ, United Kingdom
[7]Institute of Biomedical Engineering, Imperial College London, London SW7 2AZ, United Kingdom
[8]Department of Engineering Science, University of Oxford, Oxford OX1 3PJ, United Kingdom

**‡Correspondence**
Carl Emil Eskildsen,
Email: carl.eskildsen@dpag.ox.ac.uk

Molly M. Stevens,
Email: molly.stevens@dpag.ox.ac.uk

## Contents

## Supplementary Note 1: Raman model system: Methods and analysis

**Sample preparation**
Fructose (D-Fructose powder, ≥99%, Thermo Scientific) served as the analyte of interest, and sucrose (≥99.5%, Sigma-Aldrich) acted as an interferent. Stock solutions were prepared by dissolving fructose and sucrose separately in demineralized water at room temperature to final concentrations of 0.27 w/w for fructose and 0.41 w/w for sucrose. Individual samples were subsequently prepared by combining these stock solutions and additional demineralized water in varying proportions to achieve the desired analyte and interferent concentrations.

Sample preparations were carried out in 50 mL Falcon tubes. To minimize pipetting-related uncertainties, all solutions added to each sample were precisely weighed at room temperature using an analytical balance. These recorded weights were subsequently used to calculate accurate concentrations (w/w) of both analyte and interferent for each sample. An OT-2 pipetting robot (Opentrons) was then employed to transfer aliquots of the prepared samples into black, non-treated 96-well plates (Corning).

Sample compositions followed the experimental design shown in Fig. 2. Twenty-five sample compositions were defined by a $5^2$ full-factorial design with the following concentration levels for each constituent: fructose (0, 0.035, 0.070, 0.105, and 0.140 w/w) and sucrose (0, 0.055, 0.110, 0.165, and 0.220 w/w). In addition, 12 sample compositions represented analyte–interferent combinations not included in the full-factorial design.

For each of the 37 sample compositions, 32 wells were prepared in a 96-well plate, with each well representing one physical replicate sample.

Calibration, validation, and test subsets were subsequently defined computationally after spectral acquisition by randomly assigning wells from the 25 full-factorial sample compositions to calibration (2 wells per sample), validation (5 wells per sample), and test (25 wells per sample) subsets, while keeping the three repeated Raman measurements from the same well together. For the 12 additional sample compositions, 25 wells per sample were randomly selected for inclusion in the test set.

**Raman spectroscopy**
Raman spectra were acquired from each well of the 96-well plates using an in-house-built confocal Raman spectrometer.[1] The detector was cooled to -59°C. The spectrometer was configured with a magnification of 10× and a laser spot size of 80 µm. The excitation wavelength of the laser was 785 nm at a power of 30.9 mW. The integration time for each accumulation was 7.5 seconds, with a single accumulation per measurement. Three Raman spectra were recorded from each well of the 96-well plate.

**Spectral preprocessing**
Spectral preprocessing consisted of several sequential steps to ensure robust and high-quality Raman spectra for subsequent modeling. First, the spectra were truncated to include only the spectral region between 1500 $cm^{-1}$ and 400 $cm^{-1}$. Fluorescence background was then removed using asymmetric least squares smoothing, with a smoothing parameter of $10^6$, an asymmetry parameter of 0.001, and a third-order difference penalty.[2] Cosmic spikes were subsequently identified and removed through principal component analysis (PCA)-based residual detection.[3] Specifically, PCA was applied within each sample-specific spectral group, and for spectral variables with residuals exceeding five times the variable-wise standard deviation, the corresponding local spectral region (±5 variables) was replaced by PCA-estimated values (see Code Availability for a link to the GitHub repository containing source code).

Finally, second-derivative spectra were computed using the Savitzky-Golay algorithm,[4,5] applying a polynomial of order two, a window size of seven points (approx. 20 cm$^{-1}$), and calculating the second derivative. Edge effects from the derivative calculation were mitigated by removing affected spectral points, resulting in the final preprocessed spectral dataset used for modeling.

## Partial least squares regression

PLS regression projects the explanatory variables, $\mathbf{X} \in \mathbb{R}^{N \times M}$ and the response variable, $\mathbf{y} \in \mathbb{R}^{N}$ onto a shared, lower-dimensional latent space.[6] This approach is especially effective for high-dimensional data (i.e., large $M$) that exhibit significant collinearity between the explanatory variables, while maintaining a linear relationship with $\mathbf{y}$.[6] By simultaneously reducing dimensionality and maximizing the covariance between **X** and **y**, PLS often achieves robust regression performance where OLS struggles.[7]

Mathematically, PLS decomposes both **X** and **y** as follows,

$$\mathbf{X} = \mathbf{T}\mathbf{P}^{\mathrm{T}} + \mathbf{E} \qquad \text{(Supplementary Eq. 1)}$$

$$\mathbf{y} = \mathbf{T}\mathbf{q} + \mathbf{e}. \qquad \text{(Supplementary Eq. 2)}$$

In Supplementary Eq. 1, the matrix $\mathbf{T} \in \mathbb{R}^{N \times A}$ contains the latent scores, with $A$ denoting the number of LVs; $\mathbf{P} \in \mathbb{R}^{M \times A}$ represents the loadings for **X**; and $\mathbf{E} \in \mathbb{R}^{N \times M}$ represents the residuals of **X**. In Supplementary Eq. 2, $\mathbf{q} \in \mathbb{R}^{A}$ is the latent space regression coefficients; and $\mathbf{e} \in \mathbb{R}^{N}$ contains the residuals of $\mathbf{y}$.[6]

Conceptually, each LV captures a direction in which **X** and **y** exhibit maximal covariance, and the latent scores, **T**, provide a low-dimensional representation of the original data ($A < \min\{N, M\}$), effectively summarizing the key variation in **X** most strongly correlated with **y**.

The optimal number of LVs was determined by leave-one-sample-out cross-validation, balancing predictive performance against model complexity. The final PLS model was then fitted using all calibration observations. Preprocessed and mean-centred explanatory variables, pure analyte and interferent signals, and the corresponding PLS loadings are shown in Supplementary Fig. 1. The PLS model is summarised in Fig 3 and Supplementary Fig. 3, and the latent-space regression coefficients are listed in Supplementary Table 1.

## Type I and Type II error rates

Type I and Type II error rates provide complementary measures of detection performance. A Type I error (false positive) occurs when the test incorrectly declares the analyte present in a blank sample. For both samples 5 and 6 (Supplementary Fig. 4A) we collected corresponding blank samples (each with 75 observations), containing zero analyte but identical interferent levels as their respective analyte-containing samples. The LV scores of each blank observation were first projected onto the null-space (Eq. 26 in Methods, Analyte detection) to obtain hypothetical blank scores. These hypothetical blank scores were then passed through the GP models (Eqs. 11–12 and Eqs. 16–18, Methods, Gaussian-process models) to derive the observation-specific null distribution. The Type I error for each sample was computed as the fraction of blank observations whose predicted analyte values exceeded their respective detection limits calculated using Eq. 27 (Methods, Analyte detection),

$$\text{Type I error} = \frac{N_{\text{FP}}}{N_{\text{reps}}} \qquad \text{(Supplementary Eq. 3)}$$

where $N_{\text{FP}}$ is the number of false positives; and $N_{\text{reps}} = 75$ is the number of observations.

A Type II error (false negative) occurs when the test fails to detect an analyte that is in fact present. For Type II error estimation, we used 75 analyte-containing observations for each test sample (samples 5 and 6). The LV scores of each observation were first projected onto the null-space (Eq. 26 in Methods, Analyte detection) to obtain their hypothetical blank scores. These scores were passed through the GP models (Eqs. 11–12 and Eqs. 16–18, Methods, Gaussian-process models) to derive their observation-specific null distribution. Type II error was computed as the fraction of analyte-containing observations whose predicted analyte values did not exceed their respective detection limits calculated using Eq. 27 (Methods, Analyte detection),

$$\text{Type II error} = \frac{N_{\text{FN}}}{N_{\text{reps}}} \qquad \text{(Supplementary Eq. 4)}$$

where $N_{\text{FN}}$ is the number of false negatives.

By counting these false positives and false negatives separately, we obtain empirical estimates of the Type I and Type II error rates under the Gaussian-process-based and variance-only ordinary least squares-based null-distribution assumptions.

## Supplementary analyses results

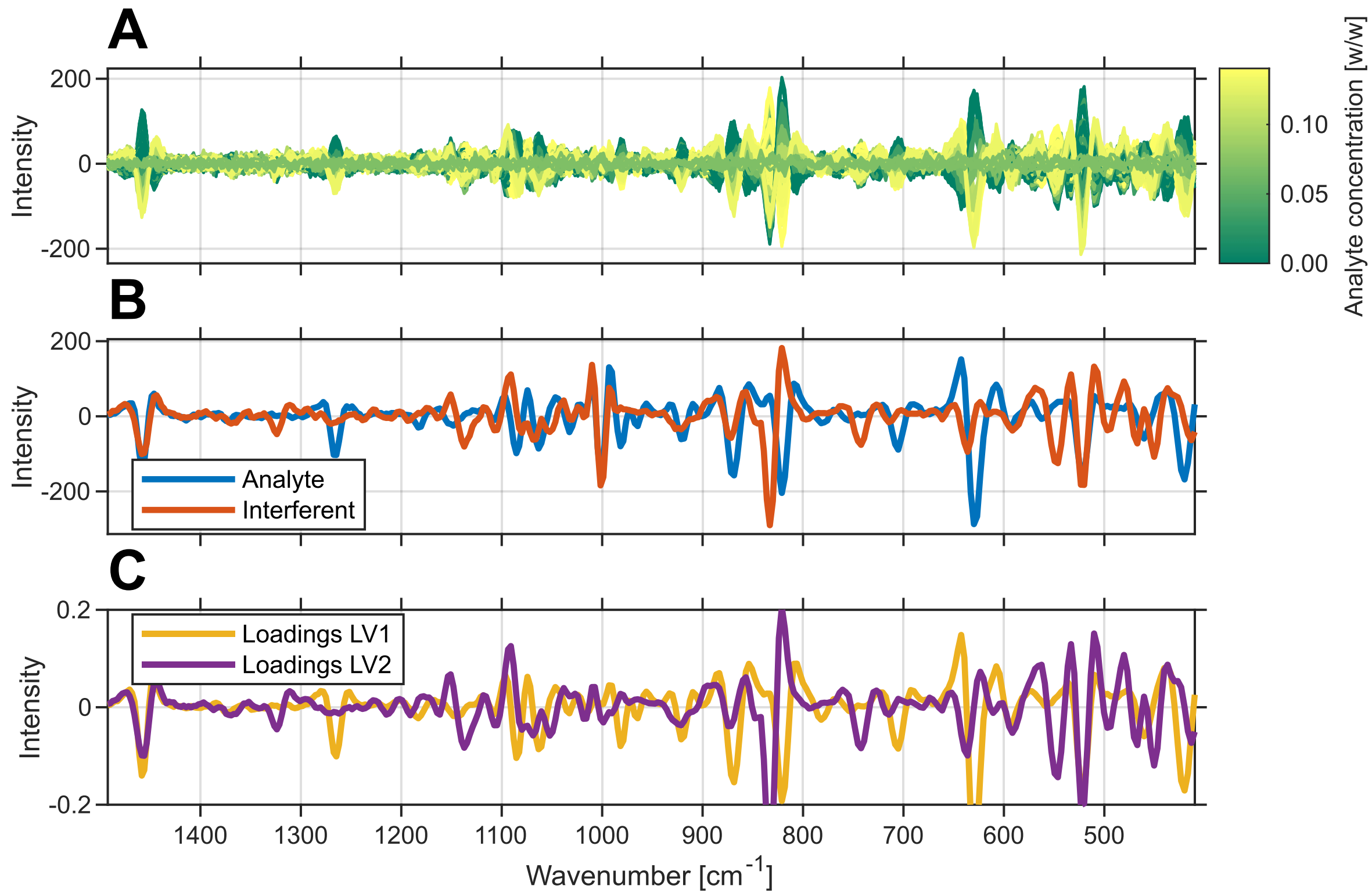

**Supplementary Fig. 1 | Explanatory variables and loadings from partial least squares (PLS) regression. A** Preprocessed and mean-centred explanatory variables (calibration samples) coloured by analyte concentration. **B** Pure analyte and interferent signals (preprocessed). **C** Loadings from PLS model. Latent variable (LV).

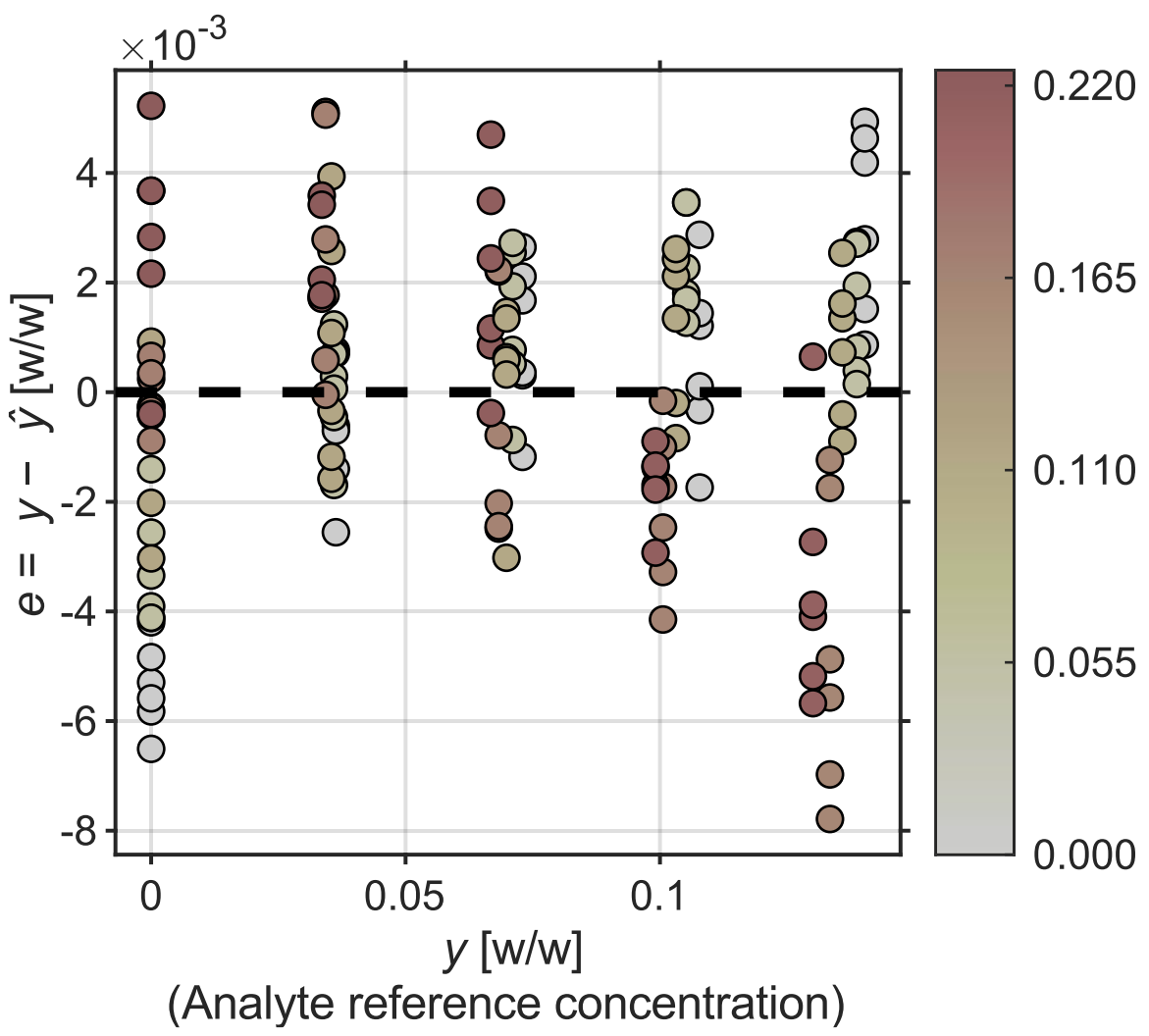

**Supplementary Fig. 2 | Calibration residuals from partial least squares (PLS) regression.** Residuals, $e = y - \hat{y}$, are plotted against analyte reference concentrations, $y$, and coloured by interferent concentration.

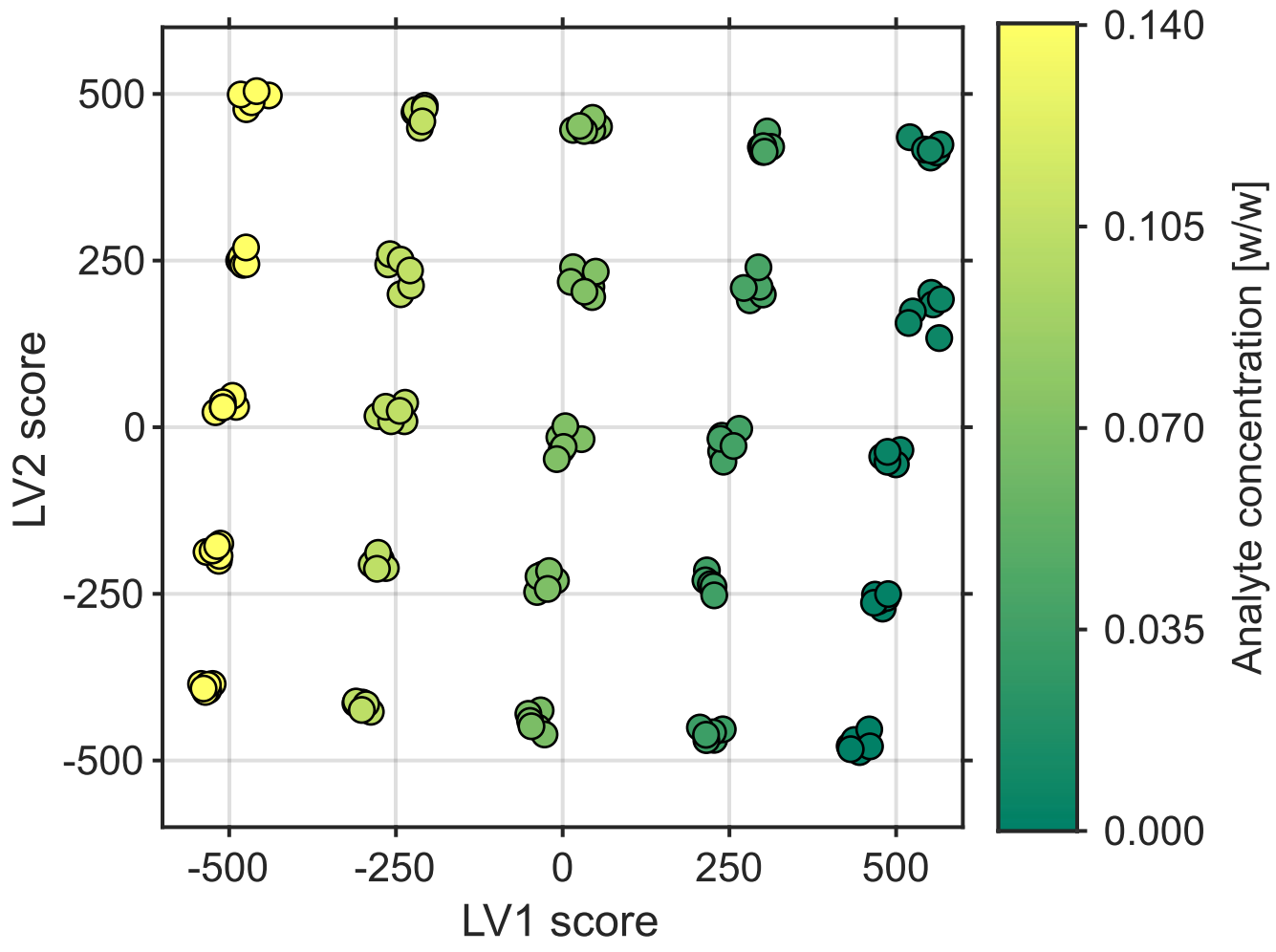

**Supplementary Fig. 3 | Partial least squares (PLS) score plot of calibration observations**, coloured by analyte concentration. Latent variable (LV).

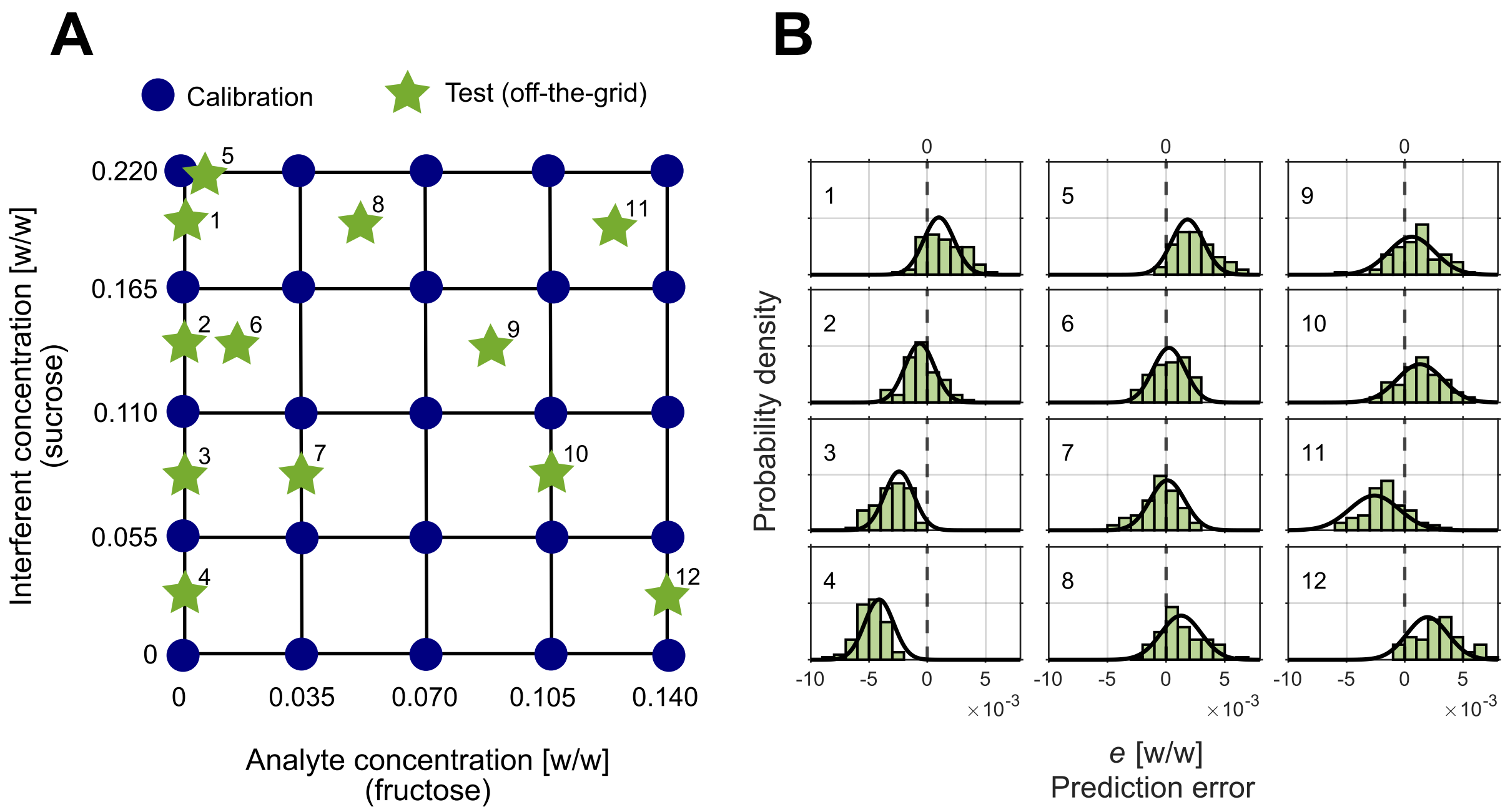

**Supplementary Fig. 4 | Prediction-error distribution for off-the-grid test samples. A** Calibration design in analyte-interferent concentration space. Blue circles show calibration mixtures and green stars show the 12 off-grid test mixtures, numbered to correspond to panel **B**. **B** Empirical distributions of PLS prediction errors for the 12 off-grid test samples (green histograms) are overlaid with the corresponding Gaussian-process (GP)-derived prediction-error distributions (black curves). Dashed vertical lines denote zero prediction error.

**Supplementary Table 1 | Latent space regression coefficients, q**, from partial least squares regression model. The coefficients $q_1$ and $q_2$ correspond to the first and second latent variables, respectively, and scale the contributions of their scores to the mean-centred analyte prediction.

| Coefficient | Value |
|---|---|
| $q_1$ | $1.330 \times 10^{-4}$ |
| $q_2$ | $-1.960 \times 10^{-5}$ |

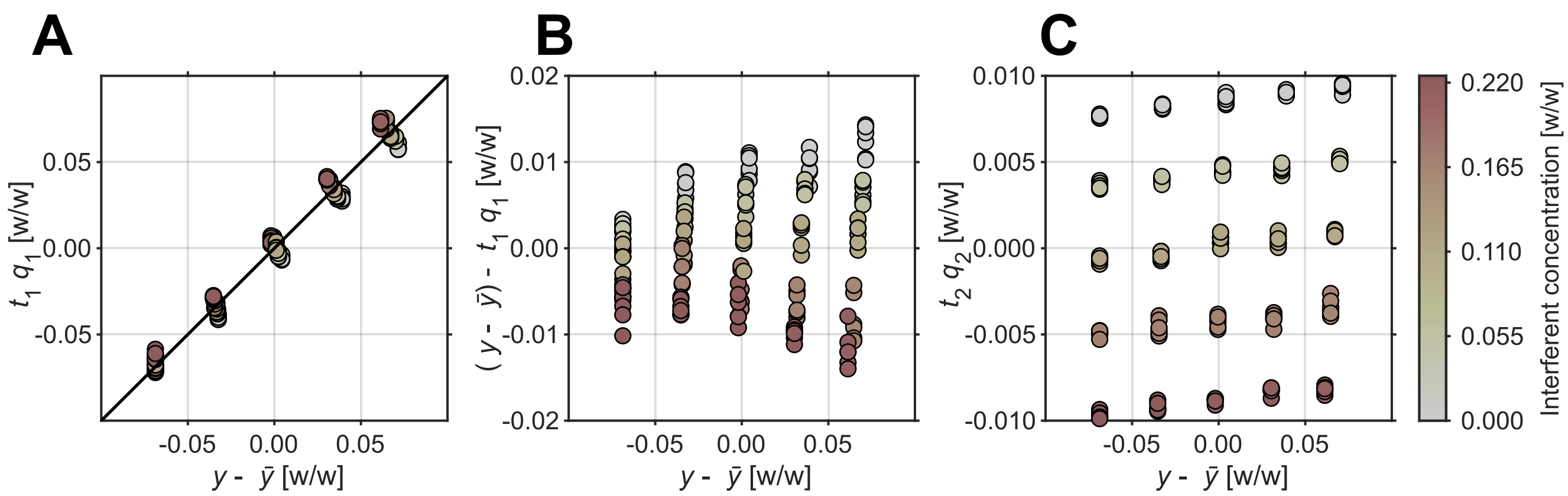

**Supplementary Fig. 5 | Latent-variable contributions to systematic prediction-error.** The figure decomposes the partial least squares prediction of mean-centred analyte concentration into contributions from the two retained latent variables. **A** Contribution from the first latent variable, $t_1 q_1$, plotted against mean-centred analyte reference concentration, $y - \bar{y}$. **B** Analyte variation remaining after removal of the first contribution, $(y - \bar{y}) - t_1 q_1$, plotted against mean-centred analyte reference concentration. **C** Contribution from the second latent variable, $t_2 q_2$, plotted against the mean-centred analyte reference concentration. The partial least squares prediction is $\hat{y} - \bar{y} = t_1 q_1 + t_2 q_2$. As these contributions vary with position in latent space, truncation of the latent representation can produce systematic prediction error whose expected value varies across that space. Points are coloured by interferent concentration.

## Supplementary Note 2: QSAR Aqueous Solubility Analysis

### Introduction

To demonstrate that the proposed uncertainty quantification framework generalises beyond spectroscopic data, we applied the method to a quantitative structure-activity relationship (QSAR) dataset. QSAR models predict molecular properties from structural descriptors and are widely used in drug discovery and environmental chemistry.[8] Unlike spectroscopic data, QSAR descriptors are heterogeneous calculated properties rather than measured intensities, providing a test of broader applicability.

### Dataset description

We used the AqSolDB database, a curated collection of experimental aqueous solubility measurements compiled by Sorkun et al. (2019).[9] The dataset contains 9,982 unique compounds with experimentally determined solubility values (LogS, mol/L) and 17 molecular descriptors calculated using RDKit: molecular weight (MolWt), octanol-water partition coefficient (MolLogP), molar refractivity (MolMR), topological polar surface area (TPSA), Labute approximate surface area (LabuteASA), counts of heavy atoms, hydrogen bond acceptors/donors, heteroatoms, rotatable bonds, valence electrons, and ring systems (aromatic, saturated, aliphatic, total), as well as the Balaban J index and Bertz complexity index. Raw data were downloaded from Harvard Dataverse[9] and processed in MATLAB for the present analysis.

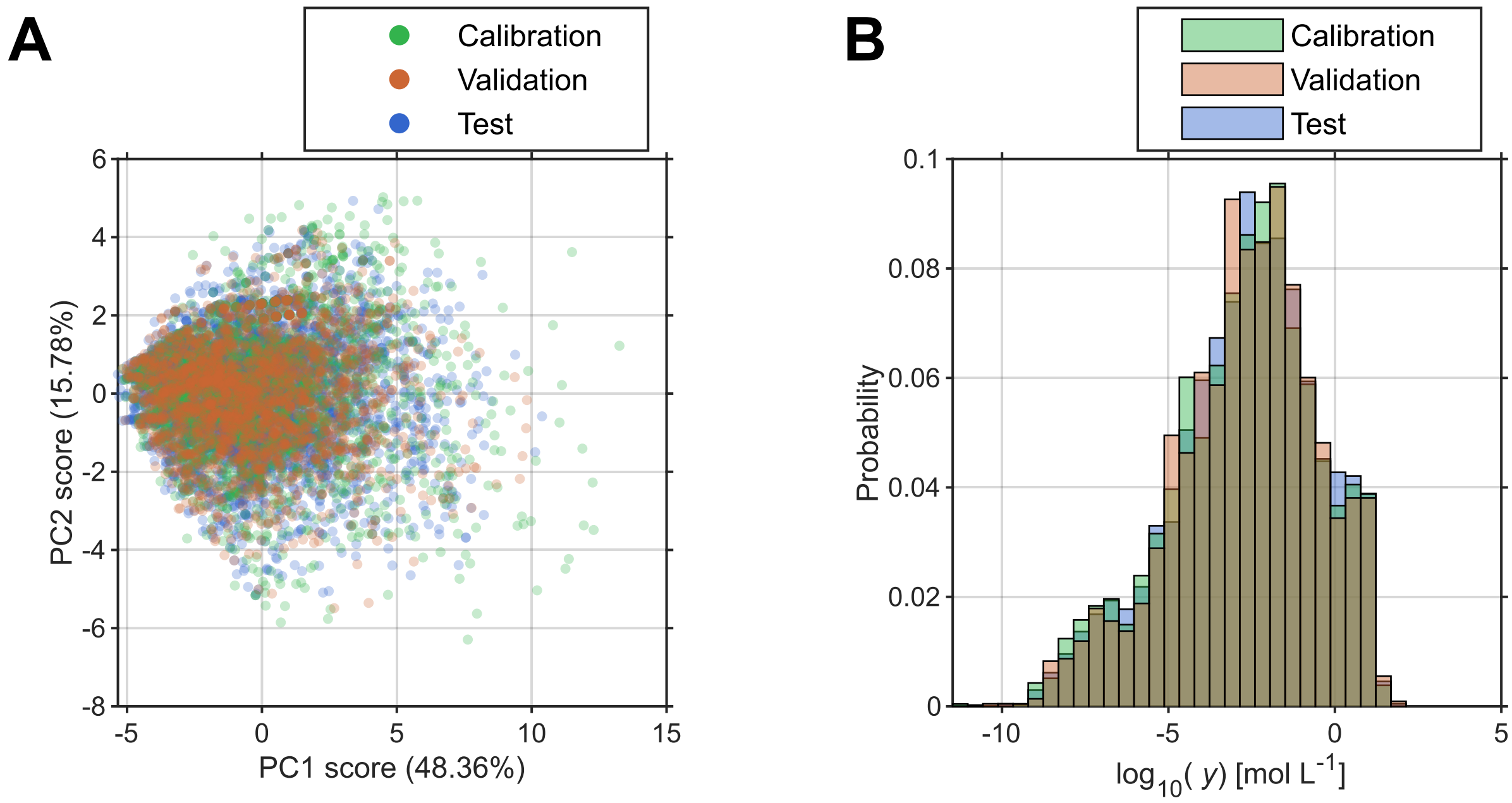

**Supplementary Fig. 6 | Overview of data split for the QSAR aqueous solubility dataset**. **A** Principal-component score plot showing the retained calibration ($n = 2{,}345$), validation ($n = 2{,}181$), and test ($n = 4{,}397$) observations. The principal component analysis model was fitted to standardized calibration descriptors after model screening; retained validation and test observations were projected into the same score space. Percentages indicate the calibration-descriptor variance explained by the first and second principal component (PC1 and PC2), respectively. **B** Distributions of the reference values, $\log_{10}(y)$, for the retained calibration, validation, and test sets, where $y$ is aqueous solubility in $\mathrm{mol\,L^{-1}}$. Histograms use common bin edges and are normalized to probability.

The data were randomly split into three sets: calibration ($25\%, n = 2{,}496$) for partial least squares (PLS) model training, validation ($25\%, n = 2{,}496$) for Gaussian process (GP) training, and test ($50\%, n = 4{,}990$) for evaluation. An initial 20-fold cross-validation on the calibration set indicated an optimal four-LV PLS model, which was used for calibration outlier screening based on absolute cross-validated response residuals exceeding three standard deviations from the median and Hotelling's $T^2$ exceeding the 99% chi-square limit. After removing calibration outliers, the PLS model was refit on the cleaned calibration set, and a second 20-fold cross-validation indicated an optimal six-LV model. A final Hotelling's $T^2$ threshold corresponding to the 99% chi-square limit of this six-LV model was then applied to the validation and test sets. This resulted in $2{,}345$ compounds for PLS calibration, $2{,}181$ compounds for GP training, and $4{,}397$ compounds for evaluation. All descriptors were autoscaled using the calibration set mean and standard deviation.

To visualise the distribution of the three data subsets, we examined both the descriptor space and the response variable. We fitted a principal component analysis (PCA) model to the calibration set (after outlier removal) and projected the validation and test sets onto the same score space (Supplementary Fig. 6A). The three subsets overlap extensively in the two-component score plot, confirming that the random split produced comparable coverage of the chemical descriptor space. Supplementary Fig. 6B shows histograms of the reference values ($\log_{10}(y)$, where $y$ is aqueous solubility in mol/L) for each subset. The response distributions likewise overlap, indicating that the three sets span a similar range of solubility values without systematic shifts. Together, these panels confirm that the random split did not introduce artefactual gaps between the sets — an important prerequisite, since the GP models are trained on the validation set and evaluated on the test set, and systematic differences between the sets would risk extrapolation artefacts in the uncertainty estimates.

**Partial least squares regression for the QSAR dataset**

A PLS model was fitted to the cleaned calibration set using 20-fold cross-validation (CV) to determine the optimal number of latent variables. Supplementary Fig. 7A displays the CV mean squared error (MSE) as a function of the number of LVs. The CV-MSE decreases up to approximately six LVs, after which the improvement plateaus. A six-LV model was therefore selected, balancing predictive accuracy against the risk of overfitting. Supplementary Fig. 7B shows the corresponding cross-validated predictions plotted against the measured log-transformed aqueous solubility values.

**Gaussian-process uncertainty models for the QSAR dataset**

Following the same two-GP framework described in the main text (Methods, Gaussian-process models), we fitted an error GP and a heteroscedastic variance GP to the validation set (n = 2,181). Both GPs used the six PLS latent scores as inputs. The error GP was fitted to the PLS prediction errors using an ARD RBF kernel (Eq. 10), estimating the spatially varying expected prediction error and epistemic variance (Eqs. 11-12). The variance GP was fitted to the residuals debiased by the error GP, using a GP prior on the log-variance function with Laplace approximation (Eqs. 13–18), estimating the spatially varying aleatoric variance. Hyperparameters for both models were optimised via negative log marginal likelihood minimisation with 10 multi-start restarts to mitigate local optima. The total prediction-error variance was obtained by summing the epistemic and aleatoric components. Both GP models were then evaluated on the held-out test set (n = 4,397).

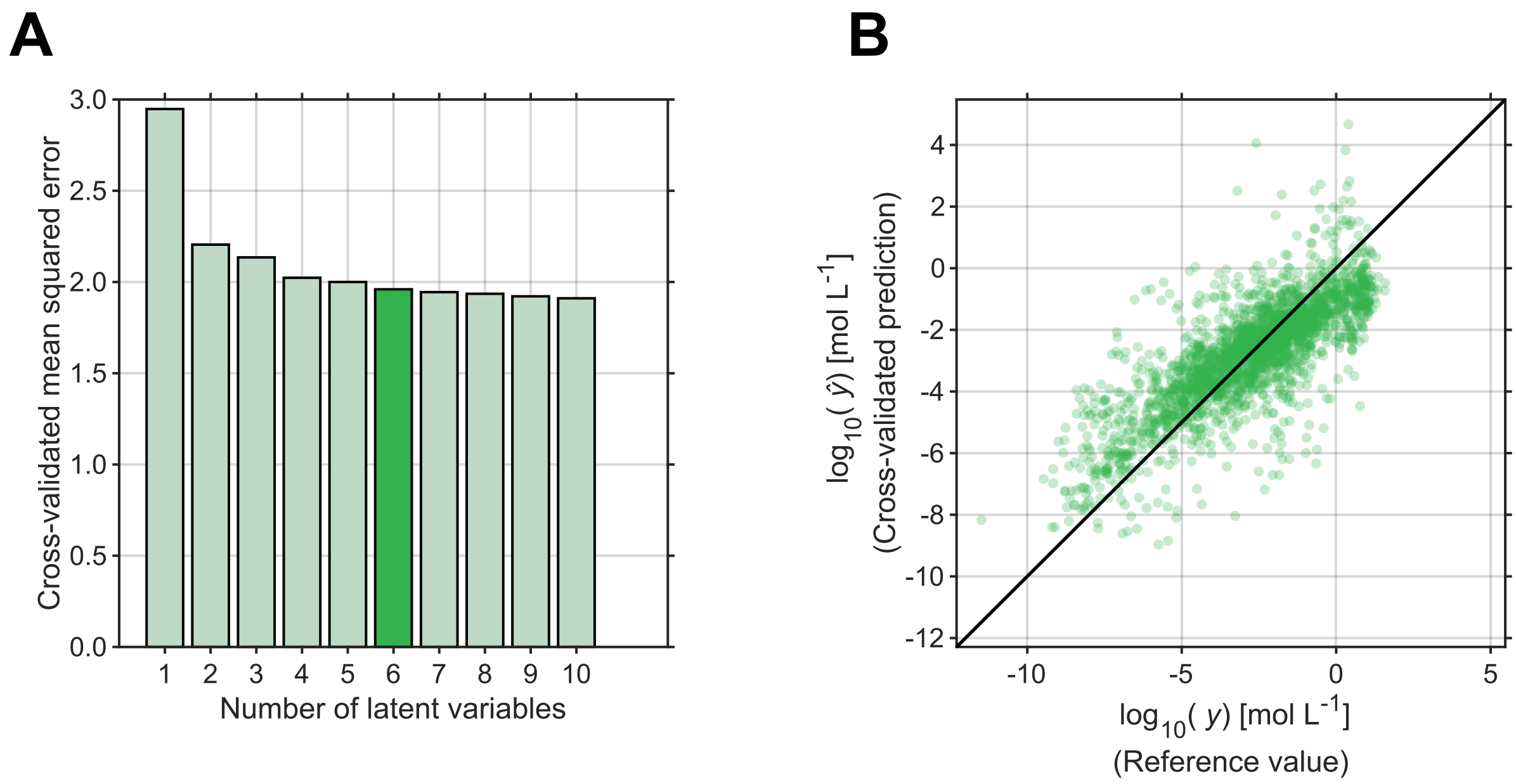

**Supplementary Fig. 7 | Calibration of the partial least squares (PLS) regression model** for the QSAR aqueous solubility dataset. **A** Cross-validated mean squared error (MSE) as a function of the number of latent variables, calculated using 20-fold cross-validation of the retained calibration set ($n = 2{,}345$). The highlighted bar indicates the selected six-latent-variable model. **B** Cross-validated predictions from the selected model plotted against the reference values, $\log_{10}(y)$, for the retained calibration observations, where $y$ denotes aqueous solubility in mol $L^{-1}$. Green points represent the cross-validated predictions; the black line indicates identity.

## Results

Supplementary Fig. 8A shows the test-set predictions before and after GP-based correction. The uncorrected PLS predictions (blue) show deviations from the identity line. After GP correction (magenta), the predictions lie somewhat closer to the identity line, consistent with an overall reduction in prediction error. Supplementary Fig. 8B quantifies this improvement: GP correction reduced the test-set MSE from 1.94 to 1.50, corresponding to a 22.7% reduction. A two-sided paired t-test comparing compound-level squared prediction errors before and after correction yielded $t(4{,}396) = 11.14$, $p < 0.0001$, and a two-sided Wilcoxon signed-rank test yielded $W = 6{,}236{,}624$, $p < 0.0001$. Because the paired squared-error differences were right-skewed and asymmetric (data not shown), these tests are interpreted as exploratory. The reduction in overall MSE is consistent with the Raman spectroscopy analysis and indicates that GP correction can improve predictions in the higher-dimensional latent space of the QSAR dataset.

Supplementary Fig. 9 summarises the uncertainty evaluation for the QSAR dataset. Supplementary Fig. 9A shows calibration plots for the GP- and OLS-derived prediction intervals, where nominal coverage is plotted against empirical coverage on the test set. For each nominal coverage level, the corresponding central Gaussian prediction interval was computed, and the empirical coverage was obtained as the fraction of test observations whose prediction errors fell within that interval. The dashed diagonal indicates perfect calibration.

As expected, the GP and OLS curves tend to converge toward the diagonal at the lower and upper ends of the coverage range, because any reasonable interval method must approach empirical coverage values of 0 and 1 as the nominal coverage approaches these limits. The key difference appears in the intermediate range: here the GP curve follows the diagonal more closely, whereas the OLS curve lies systematically above it, particularly around a nominal coverage of 0.5. This indicates that the OLS intervals are generally too wide, yielding conservative overcoverage rather than well-matched uncertainty estimates. By contrast, the GP intervals provide a closer match between nominal and empirical coverage across much of the range, indicating a more accurate uncertainty description.

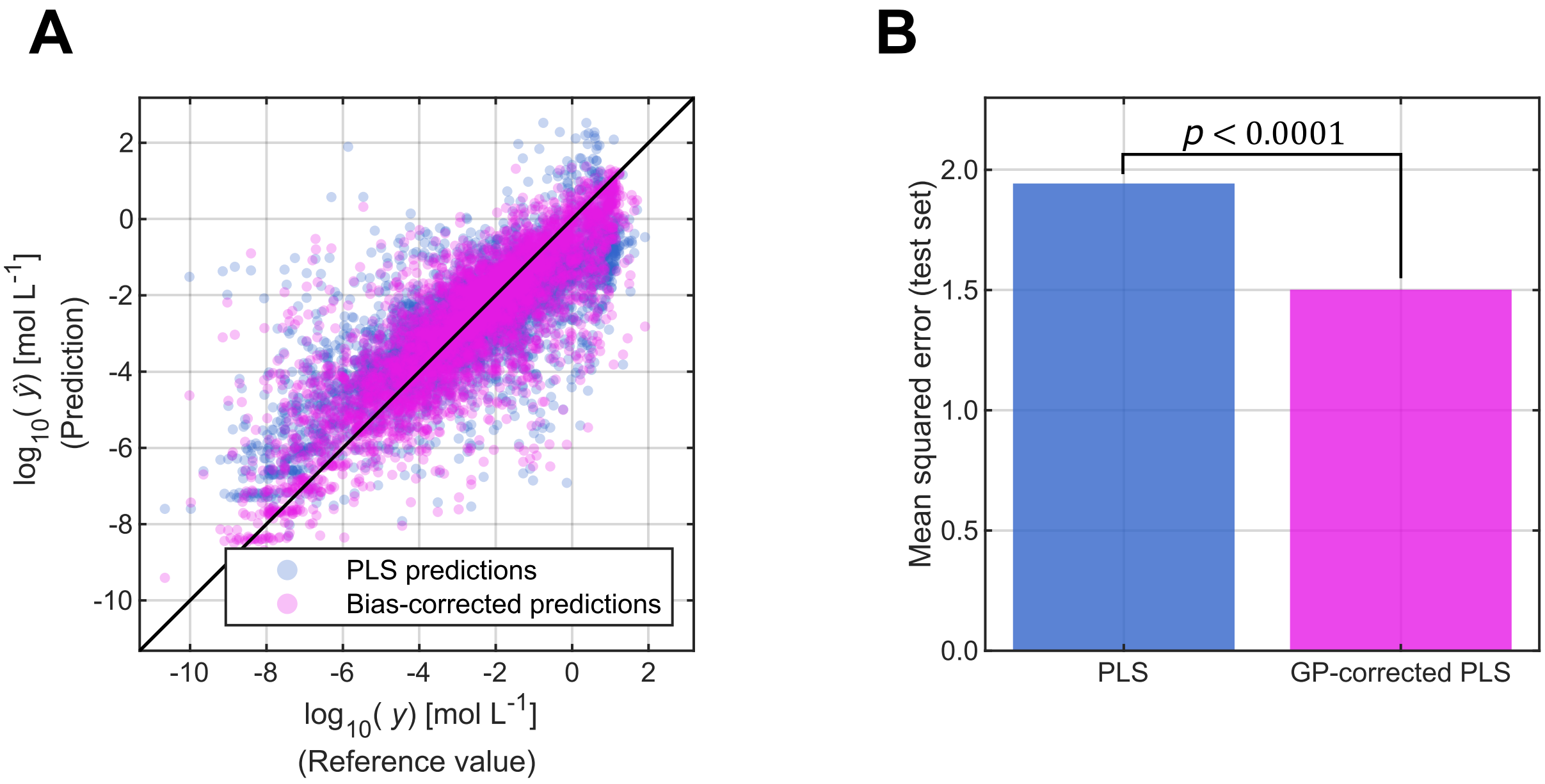

**Supplementary Fig. 8 | Effect of Gaussian-process correction on aqueous-solubility predictions. A** Uncorrected partial least squares (PLS) predictions (blue) and Gaussian-process (GP)-corrected predictions (magenta) versus reference $\log_{10}$ aqueous solubility for the retained AqSolDB test compounds. The solid black line indicates equality. **B** Mean squared error (MSE) before and after GP correction. Correction reduced the test-set MSE by 22.7%, from 1.94 to 1.50. The displayed $p$-value is from a two-sided paired t-test comparing squared prediction errors before and after correction for the same compounds ($n = 4{,}397$, $t(4{,}396) = 11.14$, $p < 0.0001$). The test is interpreted as exploratory.

Supplementary Fig. 9B explains this difference. The panel shows the distributions of the total prediction-error variance estimates for the two methods. For the GP model, the total variance is observation-specific and combines epistemic and aleatoric components, producing a broad distribution of variance estimates across the test set. In contrast, the OLS variances cluster within a very narrow range, assigning almost the same uncertainty to all compounds. Importantly, despite this flexibility, the GP model yields lower average total variance estimates than OLS, as reflected by both the mean and median variances (Supplementary Table 2). Thus, the GP achieves better overall probabilistic performance with narrower average uncertainty estimates, because it concentrates uncertainty where it is needed rather than inflating it uniformly across all compounds.

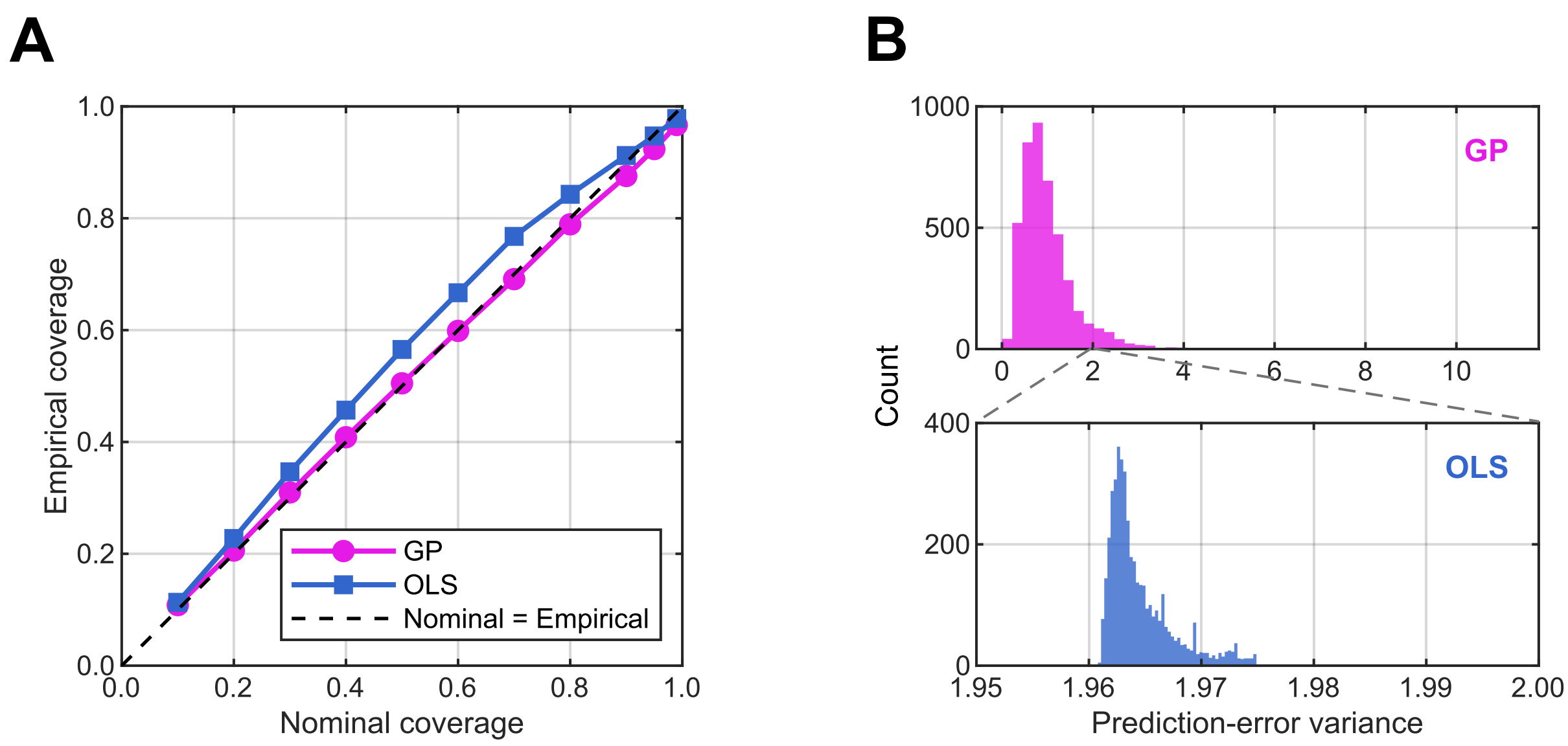

**Supplementary Fig. 9 | Evaluation of prediction-error variance estimates. A** Empirical coverage as a function of nominal coverage for Gaussian-process (GP)-derived (magenta) and ordinary least squares (OLS)-derived (blue) prediction intervals for the test set ($n = 4{,}397$). The black dashed line indicates equality between nominal and empirical coverage. **B** Distributions of total prediction-error variance estimates for GP (top; epistemic plus aleatoric variance) and OLS (bottom). Grey dashed lines delimit the $1.95 - 2.00$ range on the upper horizontal scale and connect it to the expanded scale used for the OLS histogram. The GP estimates vary broadly across compounds, whereas the OLS estimates are concentrated near $1.96$.

To summarise overall probabilistic performance, we also calculated the continuous ranked probability score (CRPS).[10] For each test observation, the CRPS compares the predicted Gaussian error distribution with the observed prediction error; lower values indicate better probabilistic predictions because both calibration and sharpness are rewarded. The GP achieved a mean CRPS of 0.62 compared with 0.76 for OLS, corresponding to a 17.8% reduction calculated using unrounded values. The corresponding summary metrics are listed in Supplementary Table 2.

**Supplementary Table 2 | Summary of uncertainty-evaluation metrics for the QSAR data.** Comparison of Gaussian process (GP)-based and ordinary least squares (OLS)-based uncertainty estimates on the test set, including mean continuous ranked probability score (CRPS) and summary statistics of the predicted total error variances. Lower CRPS indicates better overall probabilistic performance.

| | **GP approach** | **OLS approach** |
|---|---|---|
| Mean variance | 1.07 | 1.96 |
| Median variance | 0.87 | 1.96 |
| Mean CRPS | 0.62 | 0.76 |

Taken together, these results demonstrate that the GP uncertainty framework generalises beyond the spectroscopic setting of the main text. Despite the substantially different nature of the QSAR data—heterogeneous molecular descriptors rather than correlated spectral intensities, a higher-dimensional latent space (six LVs), and no replicate measurements—the error GP reduced test-set MSE by 22.7%, while the variance GP produced observation-specific uncertainty estimates that outperformed the conventional OLS approach in terms of overall probabilistic performance and did so with lower average total variance.

## Supplementary References

References in the Supplementary Information are numbered independently of the main manuscript.